\documentclass[letterpaper,twocolumn,10pt]{article}
\usepackage[english]{babel}
\usepackage{blindtext}
\pdfoutput=1

\usepackage{xspace}
\usepackage{graphicx}
\usepackage[T1]{fontenc}
\usepackage[utf8]{inputenc}
\usepackage[bottom]{footmisc}
\usepackage{usenix}
\usepackage{amsmath,amsfonts,amssymb,amsthm}
\newtheorem{theorem}{Theorem}

\newtheorem{assumption}{Assumption}

\usepackage{subfig}

\makeatletter
\let\c@subfigure\relax
\let\c@subtable\relax
\let\@listsubcaptions\relax
\let\@dottedxxxline\relax
\let\l@subfigure\relax
\let\c@lofdepth\relax
\let\l@subtable\relax
\let\c@lotdepth\relax

\let\subfloat@label\relax
\let\sf@@sub@label\relax

\makeatother
\usepackage[tight]{subfigure}

\usepackage{breakurl}           
\usepackage{url}                
\usepackage{xcolor}             
\usepackage[]{hyperref}         

\usepackage{tabularx}
\usepackage{booktabs}

\usepackage[ruled]{algorithm2e}

\usepackage{enumitem}

\usepackage{mathtools}
\usepackage{tablefootnote}
\usepackage{appendix}
\newcommand{\paraspace}{\vspace{0.05in}}
\newcommand{\parab}[1]{\paraspace\noindent{\bf #1}}

\newcommand{\sys}{{DLCP}\xspace}

\newenvironment{icompact}{
\begin{itemize}[topsep=2.5pt, partopsep=0pt, leftmargin=1em]
  \setlength{\itemsep}{2.5pt}
  \setlength{\parskip}{0pt}
  \setlength{\parsep}{0pt}
}{\end{itemize}}

\newcounter{packednmbr}

\newcommand{\eg}{{\it e.g.,}\xspace}

\title{Domain-specific Communication Optimization for Distributed DNN Training}
\author{Hao Wang$^1$, Jingrong Chen$^1$, Xinchen Wan$^1$, Han Tian$^1$, Jiacheng Xia$^1$, Gaoxiong Zeng$^1$, Weiyan Wang$^1$\\ Kai Chen$^1$, Wei Bai$^2$, Junchen Jiang$^{3}$\\
$^1$iSING Lab, Hong Kong University of Science and Technology\\
$^2$Microsoft Research, $^3$University of Chicago}

\begin{document}
\pagestyle{plain} 



\maketitle
\begin{abstract}
Communication overhead poses an important obstacle to distributed DNN training and draws increasing attention in recent years. Despite continuous efforts, prior solutions such as gradient compression/reduction, compute/communication overlapping and layer-wise flow scheduling, etc., are still coarse-grained and insufficient for an efficient distributed training especially when the network is under pressure. 

We present \sys, a novel solution exploiting the domain-specific properties of deep learning to optimize communication overhead of DNN training in a fine-grained manner. At its heart, \sys comprises of several key innovations beyond prior work: e.g., it exploits {\em bounded loss tolerance} of SGD-based training to improve tail communication latency which cannot be avoided purely through gradient compression. It then performs fine-grained packet-level prioritization and dropping, as opposed to flow-level scheduling, based on layers and magnitudes of gradients to further speedup model convergence without affecting accuracy. In addition, it leverages inter-packet order-independency to perform per-packet load balancing without causing classical re-ordering issues. \sys works with both Parameter Server and collective communication routines. We have implemented \sys with commodity switches, integrated it with various training frameworks including TensorFlow, MXNet and PyTorch, and deployed it in our small-scale testbed with 10 Nvidia V100 GPUs. Our testbed experiments and large-scale simulations show that \sys delivers up to $84.3\%$ additional training acceleration over the best existing solutions.


\end{abstract}
\section{Introduction}

Deep Learning (DL) plays a key role in modern AI applications such as Computer Vision\cite{resnet2016cvpr, Alexnet, VGG} and Natural Language Processing\cite{wu2016google, venugopalan2015sequence, merity2017regularizing} etc. At the core of DL, however, training Deep Neural Networks (DNNs) can be notoriously time-consuming, due primarily to sheer volumes of data communication and growing model complexities. Although computation (forward/backward propagation) can be parallelized via mini-batching, DNN training still has 100s of iterations, each of which ends with exchanges of massive gradient updates across 10s to 100s of distributed workers, potentially causing severe worst-case congestion and tail latencies and slowing DNN training down to a crawl. These communication bottlenecks have been reported in real production~\cite{bytescheduler} and in recent literature~\cite{p3,tictac,pipedream,parameterhub, rat-apnet}.


Substantial efforts are made recently to alleviate the communication bottleneck in DNN training. For instance, there are techniques to compress/reduce the amount of gradient updates by exploiting the sparsity of gradient values (\eg mostly zeros)~\cite{sparse2018nips, deep2017arxiv} or SGD's inherent tolerance to asynchronous gradient updates~\cite{gaia-nsdi,ssp}. Others seek to mitigate communication overhead by pipelining communication and computation through careful layer-wise scheduling (\eg forward propagation computes the gradients of front layers while synchronizing deeper layers gradients)~\cite{tictac,bytescheduler,gpipe}.

While these solutions exploit DL-specific properties, they operate on the application layer and can nevertheless suffer from poor {\em tail} performance when the (datacenter) network is under stress. For example, although gradient compression reduces the overall traffic volume, but they are not immune to long-tail delays caused by transient packet drops or queueing resulting from bursty traffic spikes (see $\S$\ref{subsec:problem}). Likewise, even if a DNN scheduler prioritizes certain flows (\eg a much needed tensors), it only controls when end-hosts initiate these flows and switches {\em in network} will ignore these application-specific priorities when choosing which packets to be queued or dropped (e.g.,~\cite{p3, bytescheduler}). In other words, most of these optimizations operate at the flow-level (\eg each flow is a tensor~\cite{ps2014osdi} or tensor partition~\cite{p3, bytescheduler}) at the end-host, thus insufficient to handle packet-level hiccups in the network (e.g., packet queueing or dropping ~\cite{bai2014pias, zhang2016guaranteeing}). 


In this paper, we argue that it is fundamentally more effective to embrace these domain-specific properties of DL at lower layers of the network stack. 
Such a ``holistic'' approach can potentially enable similar DL optimizations and more ($\S$\ref{subsec:opportunities}) at a much finer {\em per-packet} granularity, whereas prior work is restricted to {\em flow}-level optimizations.
For instance, transport layer can avoid tail flow latency by intentionally ignoring the small fraction of data delayed or missed by the network.~\cite{xia2019rethinking} Switches in the network can prioritize packets that carry important gradients to respect the application-specific semantics. Furthermore, the independence among packets in DNN training allows for per-packet load balancing without the need for packet reordering.

We present \sys, a novel solution exploiting the domain-specific properties of deep learning to optimize communication overhead of DNN training in a fine-grained manner. 
One of the key enabling concepts of \sys is {\em bounded loss tolerance}. 
Reliability in current transport control is ``all-or-nothing'': TCP requires all packets to be received and can be blocked by a tiny fraction of delayed packets; whereas UDP has no reliability guarantee at all. 
In contrast, what matters to DNN training is a certain fraction ($<100\%$) of data, {\em not} all data, are received, due to the SGD-based training ($\S$\ref{subsec:opportunities}). 
\sys exploits such domain-specific insight to design a simple yet effective bounded-loss tolerant end-host transport that minimizes the gradient transmission time, by intentionally ignoring packets (bounded by $p$) delayed or lost in the network without retransmission ($\S$\ref{sec:design}). Our result shows that this effectively cuts tail latency which cannot be avoided purely through gradient compression ($\S\ref{subsec:problem}$).


Another key novelty behind \sys is to set packet priorities based on layers and magnitudes of gradients, and enforce prioritized queueing or dropping in the switch that matches DNN training semantics.
Our insight is that gradients can be different in two dimensions ($\S$\ref{subsec:opportunities}). On one hand, gradients of front layers are more loss tolerant than back layers in terms of model convergence. This is because, in deep neural networks, different layers extract features in different levels of abstraction. Back layers generally contain accumulated information that is learned based upon information in front layers, so more important~\cite{yosinski2014transferable, lecun2015deep}.
On the other hand, gradients of larger magnitude have more impact and are less loss tolerant, as their losses can negatively affect the convergence. This is because, during training, SGD leverages gradients to learn the correlations between the intermediate features and the model output. For a given dataset, larger gradients possess stronger correlations between the connected features and the task than of small gradients, and thus are more important. \sys leverages such observations to innovate switch mechanisms. 


In addition, \sys leverages inter-packet order-independency to enforce per-packet load balancing in the network without raising classical re-ordering concerns. The insight behind this is, unlike traditional applications where a message usually contains multiple packets (thus  order-dependent), a packet in DNN training consists of multiple messages (gradients or parameters), thus inter-packet has no ordering ($\S$\ref{subsec:opportunities}). This enables ideal per-packet load balancing to fully utilizes network bandwidth.  

\sys works with both Parameter Server and collective communication routines and supports various training frameworks such as TensorFlow, MXNet and PyTorch, etc. We have implemented \sys with commodity switches, integrated it with all the three training frameworks mentioned above, and deployed it over our small-scale testbed with 10x Nvidia V100 GPUs ($\S$\ref{sec:impl}). Through testbed experiments and large-scale simulations, we found that ($\S$\ref{sec:eval}):
\begin{icompact}
\item Compared to prior optimization schemes such as P3~\cite{p3} and ByteScheduler~\cite{bytescheduler}, \sys delivers up to $84.3\%$ additional training speedup across different DNN models, due to its in-depth domain-specific optimizations. 

\item \sys achieves performance improvement in both PS and Ring All-Reduce, for PS the speedup is $84.3\%$, and for Ring All-Reduce the value is $11\%$.

\item Compared to general datacenter transport solutions such as DCTCP~\cite{dctcp} and pFabric~\cite{pfabric2013sigcomm}, \sys provides up to 186\% better FCT under pressing traffic due to its bounded-loss tolerance and per-packet load balancing. 


	
\end{icompact}



\section{Background and Motivation}\label{sec:motivation}


\begin{figure}
    \centering
    \includegraphics[width=0.7\linewidth]{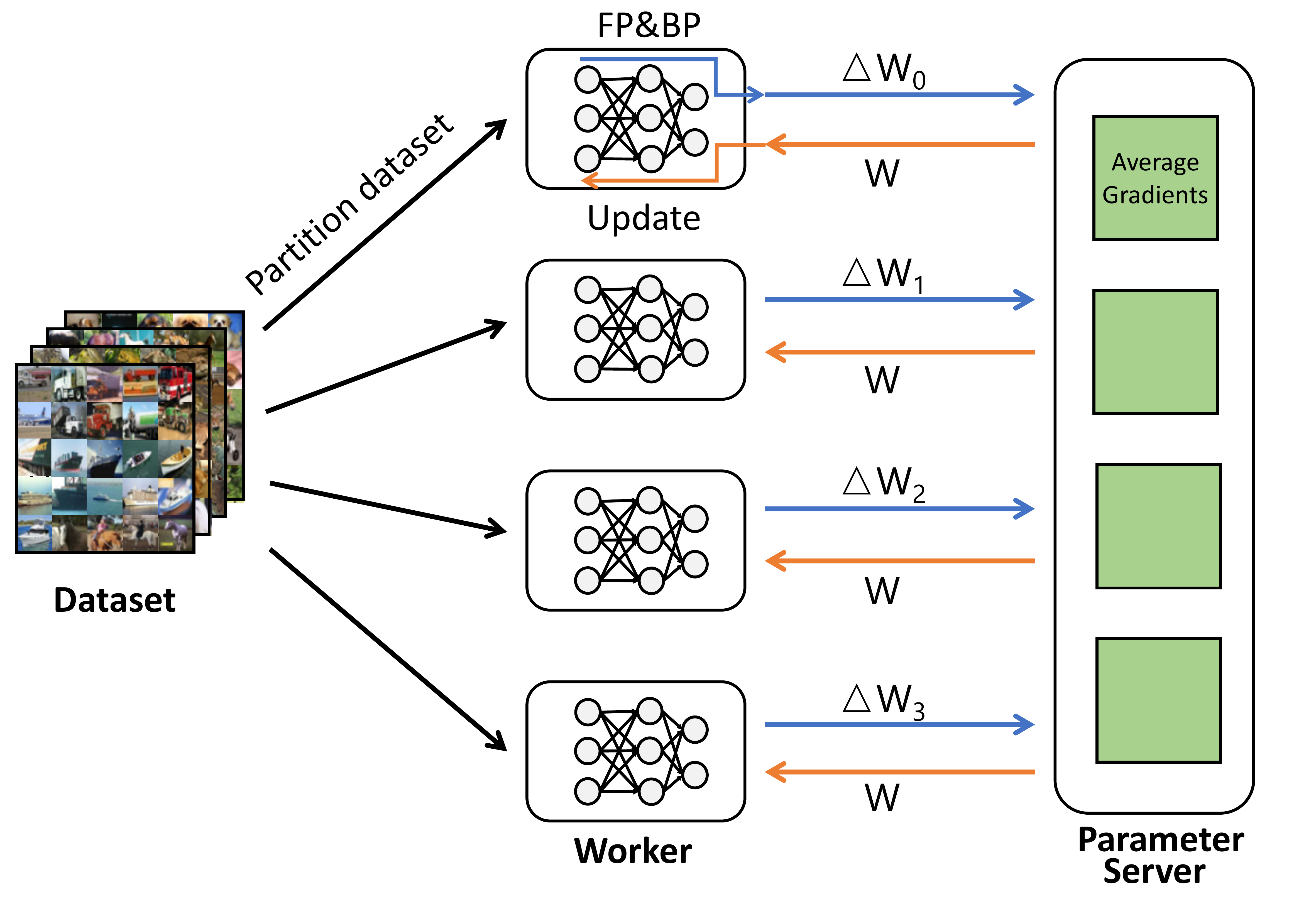}
    \caption{Data Parallelism (Parameter Server) }
    \label{fig:distml}
\end{figure}

\subsection{Distributed DNN Training}\label{subsec:training}

The goal of DNN training is to learn intricate representations of large datasets with multiple levels of abstraction~\cite{lecun2015deep}.
For this purpose, a DNN model is iterated by a large dataset many times (or ``epochs'') to minimize a loss function. 

\parab{Training process:} An epoch of DNN training consists of multiple iterations. The procedure of an iteration is typically as follows: (1) The DNN model and a partition (or ``mini-batch'') of data are taken as input; (2) the mini-batch travels through the model from the first layer to the last layer and computes the model loss, which is termed \textit{forward propagation (FP)}; (3) with the model loss derived, it computes the gradients backwards from the last layer to the first layer, which is called \textit{backward propagation (BP)}; and finally (4) the gradients, which represent information acquired from the mini-batch, are used to update the model with some optimization algorithms \eg Stochastic Gradient Descent (SGD)~\cite{SGDopt,Adamopt,Momentumopt}. After that, the training proceeds to the next iteration. 


\parab{Data parallelism:} DNN training is often time-consuming with complex models and large datasets, e.g., training of ResNet-50 with ImageNet~\cite{imagenet} costs 29 hours on 8 Tesla P100 GPUs~\cite{resnet2016cvpr} and 115.2 minutes on 8 latest V100 GPUs~\cite{mlperf}, respectively. To speed up, current practice is to leverage data parallelism\cite{chilimbi2014project,dean2012large,krizhevsky2014one}, in which mini-batches of training data are distributed across multiple machines (or ``workers'') as shown in Figure~\ref{fig:distml}. Different workers share the same global model, and compute gradients (with FP and BP) using their respective mini-batches individually. Then, gradients from all workers are synchronized and aggregated to update the global model\footnote{Generally, there are three approaches for model synchronization: Bulk Synchronous Parallelism (BSP)\cite{bsp}, Asynchronous Parallel (ASP)~\cite{hogwild2011nips}, and Stale Synchronous Parallel (SSP)~\cite{petuum}. Among them, BSP, in which all workers need to trained on the same iterations, is typically adopted in production\cite{sgd1hour, pipedream ,tensorflow}. This is because, compared to ASP or SSP, BSP has its simplicity and good convergence performance~\cite{tensorflow,revisiting}. Furthermore, BSP produces deterministic and reproducible results, which give it a great advantage for hyper-parameter tuning \cite{gandiva}. We assume BSP throughout this paper.}, using either parameter server architecture\cite{ps2014osdi} or collective routines like all-reduce\cite{allreduce}:
\begin{icompact}
    \item \textit{Parameter server (PS)}~\cite{ps2014osdi} is a logically centralized key-value store. In each iteration, workers pull new model parameters from PS for training, and then push gradients to PS for model updating. PS enables flexible parameter synchronization pattern and is generally fault-tolerant. 
    \item \textit{All-reduce}~\cite{allreduce} is a collective operation to sum up gradients of all workers. A popular implementation is ring-allreduce~\cite{horovod}, in which workers form a logical ring and each worker has two neighbors. During gradient aggregation, each worker receives a chunk of gradients from its left neighbor, add to its local copy, and send the chunk to the right neighbor, until all gradients are updated. Compared to PS, ring-allreduce generates more uniform traffic pattern, but is error-prone due to long communication channel.    
\end{icompact}

\parab{Communication bottleneck:} The network communication is heavily involved in the above model synchronization~\cite{bytescheduler}. Within each iteration, each worker may need to send and receive model gradients/parameters with tens to thousands of MBs \cite{resnet2016cvpr, VGG, Alexnet} at milliseconds. Consequently, communication often consumes a significant amount of the total training time, and poses an important bottleneck in distributed DNN training. This phenomenon has been observed by tons of recent literatures \cite{pipedream,aluminum,network-centric,ps2014osdi,poseidon,bytescheduler,p3,tictac}. For example, training AlexNet on 8 nodes demands more than 26Gbps bandwidth to avoid blocking\cite{poseidon}. Furthermore, a recent measurement has shown that communication accounts for as high as $90\%$ of total training time over 32 GPUs \cite{pipedream}. Worse, as reported by a large online service provider, due to communication overhead, the training performance is far from linear speed-up with an increasing number of GPU servers in many of their both internal and publicly available training workloads~\cite{bytescheduler}.  

\begin{table*}[t]
\centering
\small
\begin{tabular}{|c|c|c|c|c|}
\hline
Bounded loss ratio ($p$) & \multicolumn{2}{c|}{0\%-1\%} & 1\%-2\% & >2\% \\ 
\hline
\hline
        & \multicolumn{1}{l}{LSTM \cite{lstm} (0.6\%)} & \multicolumn{1}{l|}{VGG16 \cite{VGG} (0.7\%)}          & \multicolumn{1}{c|}{ResNet34 \cite{resnet2016cvpr} (1\%)}   & \multicolumn{1}{c|}{Wide ResNet50 \cite{WRN} (2.5\%)} \\ 
 
Model    & \multicolumn{1}{l}{AlexNet \cite{Alexnet}(0.8\%)} & \multicolumn{1}{l|}{VGG13 \cite{VGG} (0.9\%)}     & \multicolumn{1}{c|}{GRU \cite{gru} (1.2\%)}   & \multicolumn{1}{c|}{ResNet50 \cite{resnet2016cvpr} (2.5\%)}   \\  
 
         & \multicolumn{1}{l}{ResNet18 \cite{resnet2016cvpr} (0.9\%)} & \multicolumn{1}{l|}{VGG19 \cite{VGG} (0.9\%)} & \multicolumn{1}{c|}{Wide ResNet101 \cite{WRN} (1.3\%)}     & \multicolumn{1}{c|}{ResNet101 \cite{resnet2016cvpr} (3.5\%)} \\  
\hline
\end{tabular}
\caption{The  bounded  loss  tolerance  across  a  wide  range  of  DNN  models. For LSTM and GRU models, we train them as NLP tasks using wikitext-2\cite{wikitext}. And other models are trained as CV tasks using Caltech101\cite{caltech101}.
}
\label{table:bounded_loss}
\vspace{-4mm}
\end{table*}

\subsection{Existing Solutions and Problems}\label{subsec:problem}
To overcome the communication bottleneck, many solutions~\cite{poseidon, deep2017arxiv, ssp, pipedream, bytescheduler, allreduce} have been proposed recently. Among them, we review two lines of solutions that are closely related to us, point out their problems, which motivate our work. We leave the discussion of other related work to $\S$\ref{sec:related}. 


\begin{figure}[htbp]
    \centering
    \includegraphics[width=.8\linewidth]{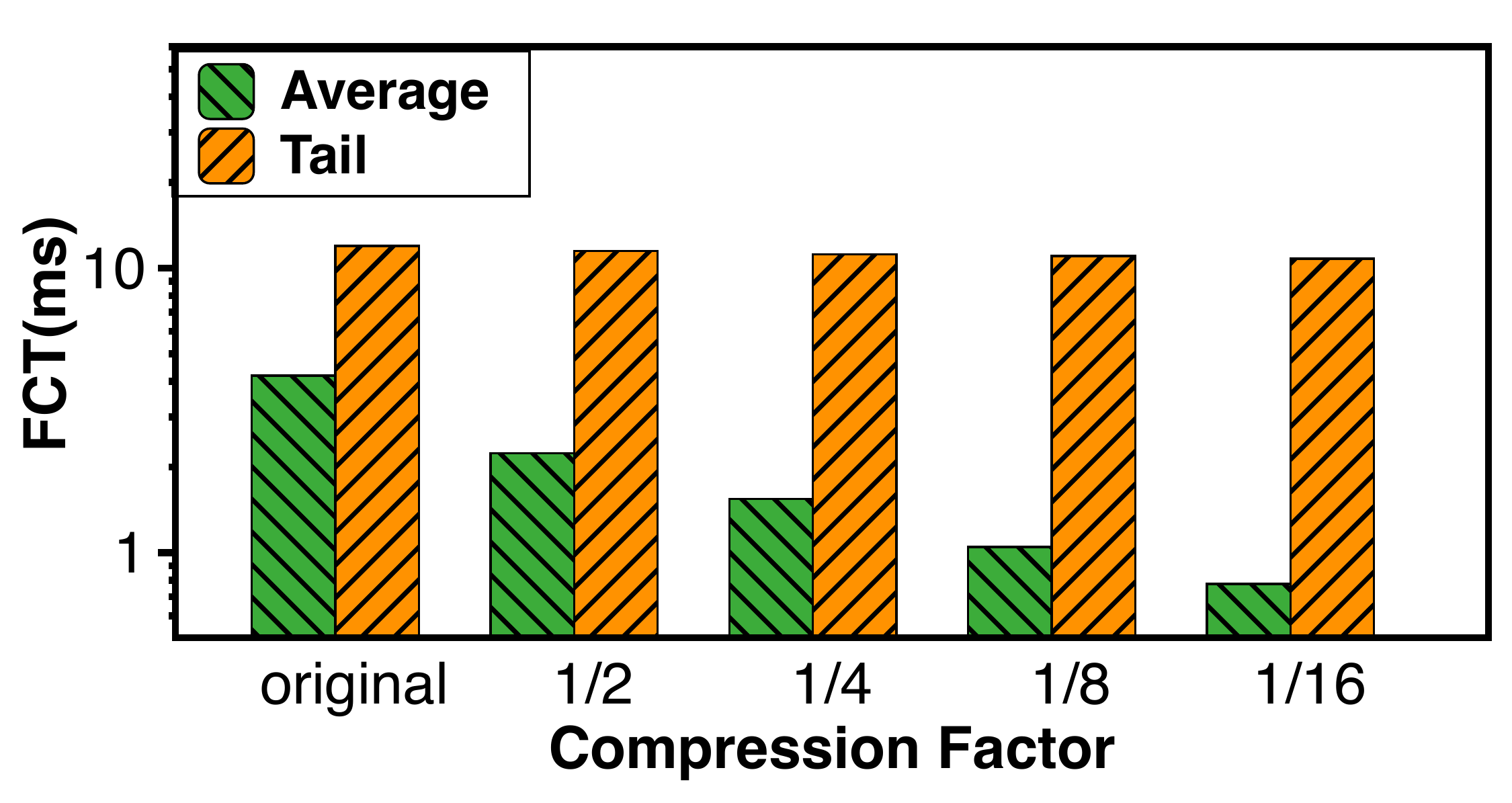}
    \caption{The average and tail FCTs under different gradient compression factors, by simulating communication pattern of training GoogleNet\cite{googlenet} in PS. 
    }
    \label{fig:motiv:reduce_size_tail}
\end{figure}

\parab{Gradient sparsification \& quantization:} One line of work proposed to reduce the network load of distributed DNN training through gradient sparsification~\cite{deep2017arxiv, sparse2018nips} or quantization~\cite{qsgd2017nips}. Specifically, gradient sparsification reduces network traffic by applying a filter and only sends gradients of large magnitude, whereas gradient quantization represents the gradients with lower-precision floating point numbers to reduce traffic volume. While both approaches help reduce overall network traffic, they do not make communication completely immune to long-tail latencies due to transient packet drops or queueing, either self-inflicted or by competing applications. The key reason is that the tail latency is often caused by the communication pattern, not only the traffic volume.

We used experiment to demonstrate the problem. In our experiment, we use ns3 to simulate the communication process of training GoogleNet, a widely used model, in PS\cite{ps2014osdi} over 80 workers (colocated with 80 servers) under the same rank, the parameters are equally divided, and the total parameter size of GoogleNet is 6.8M. The bandwidth is 100Gbps, the switch buffer size is 16MB, fetching from most commodity switches, we use TCP NewReno\cite{tcpnewreno} and RTOmin is 10ms, a commonly used setting\cite{rtomin10ms}. We compare two cases: (1) original training job, and (2) training job with gradient compression reducing traffic volume down to from 1/2 to 1/16 of the original size. Figure~\ref{fig:motiv:reduce_size_tail} shows the results of both average and tail FCTs. As we can see, while the gradient compression  does reduce the average FCT steadily (from 4.19 ms to 0.78 ms) as compression factor decreases (from 1/2 to 1/16), the tail FCT, which eventually decides the overall communication time of one training iteration remains almost the same around 10 ms. Such long-tail latency compromises the overall training efficiency, undermining the benefit brought by gradient sparsification or quantization.



\begin{figure*}[htbp!]
    \centering
    \subfigure[Baseline]{
	\includegraphics[height=.135\linewidth]{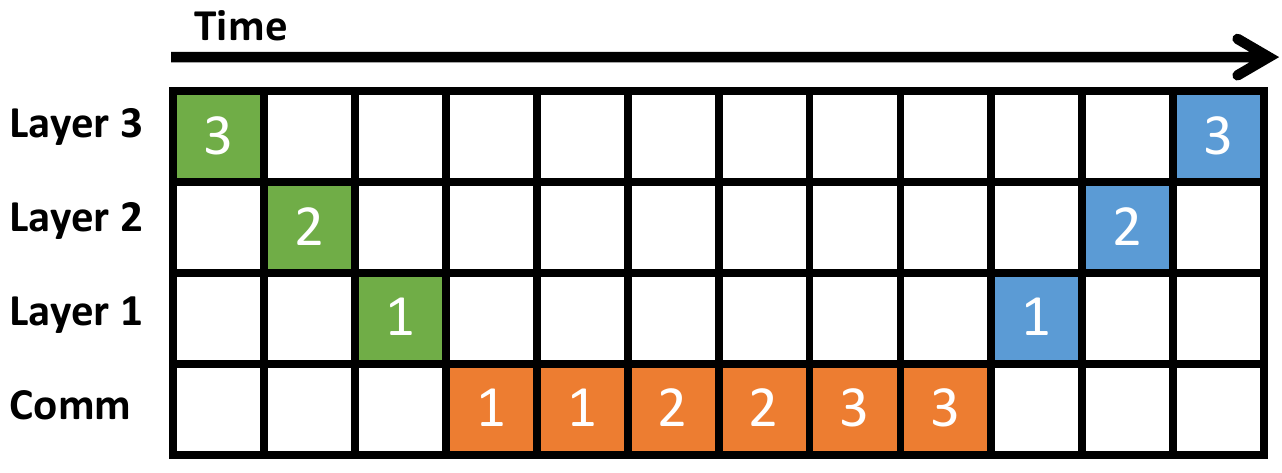}
    }
    \subfigure[Communication/computation overlapping]{
    	\includegraphics[height=.135\linewidth]{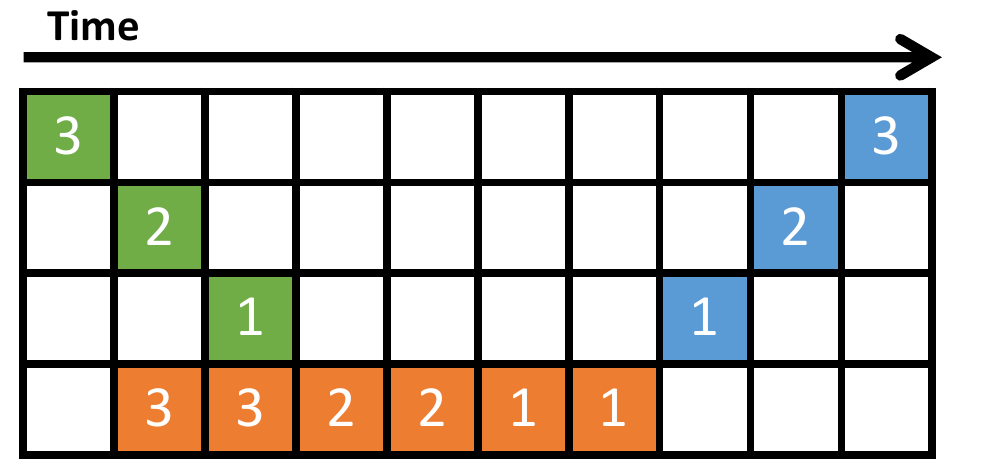}
    }
    \subfigure[Priority-based scheduling]{
	\includegraphics[height=.135\linewidth]{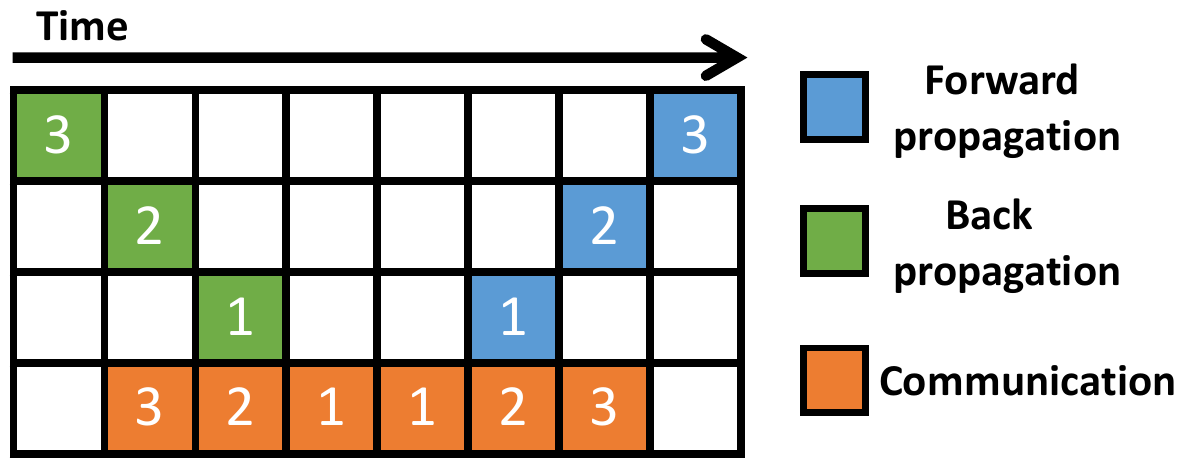}
    }  
    \caption{Optimization by layer-wise scheduling. We assume communication and computation, FP and BP take the same time. }
    \label{fig:background:scheduling}
\end{figure*}

\parab{Communication/computation overlapping \& scheduling:} Another line of work optimizes the overall training performance by overlapping  communication and computation. In this thread, Poseidon~\cite{poseidon}, together with some training frameworks like MXNet, PyTorch and TensorFlow, take the initiative to overlap communication and computation (more specifically, BP). The insight behind these solutions is the layer-by-layer DNN structure and the independence between gradient communication of one layer and gradient computation of another. Instead of waiting for the completion of the entire BP, they transmit gradients of a layer once they are ready, parallelizing the gradient communication of this layer with gradient computation of other layers (Figure~\ref{fig:background:scheduling}(b))  

On top of Poseidon~\cite{poseidon}, P3~\cite{p3}, TicTac~\cite{tictac} and Bytescheduler~\cite{bytescheduler} move one step further to overlap communication of the current iteration with FP of the next iteration, by tensor partitioning and priority-based scheduling. The insight behind these approaches is the order of the gradients/parameters consumed in the subsequent training iteration. As shown in Figure~\ref{fig:background:scheduling}(c), they prioritize transmission of layer $i$ data over that of layer $j$ (for $i<j$), which potentially accelerates the training pipelining by starting the next iteration FP earlier. 

While promising, these solutions are still insufficient, because: (1) they do not directly solve the tail latency issues introduced above; and (2) they are all purely end-host based solutions which control how end-hosts schedule these flows, but network switches are unaware of such application-specific priorities when choosing which packets to be dequeued. In other words, they only schedule at flow-level at the end, thus unable to handle packet-level hiccups in the network.


\subsection{Observations and New Opportunities}\label{subsec:opportunities}
By exploiting the domain-specific properties of DNN training, we make the following key observations which provide new opportunities for  communication optimization, addressing the above problems. 


\begin{figure}[h!]
\subfigure[RNN]{
\begin{minipage}[b]{0.225\textwidth}
\includegraphics[width=\textwidth]{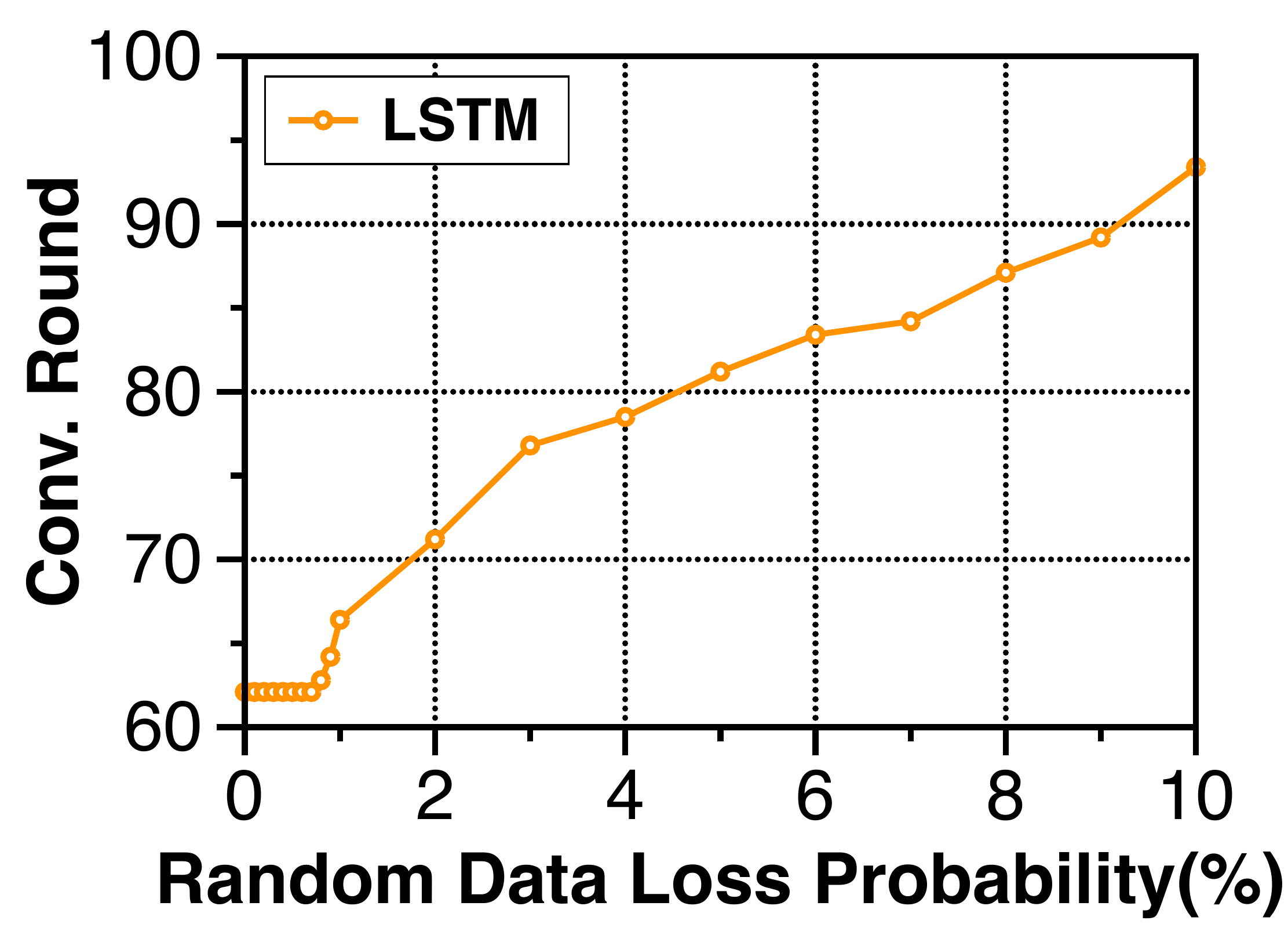}
\end{minipage}
}
\subfigure[CNN]{
\begin{minipage}[b]{0.225\textwidth}
\includegraphics[width=\textwidth]{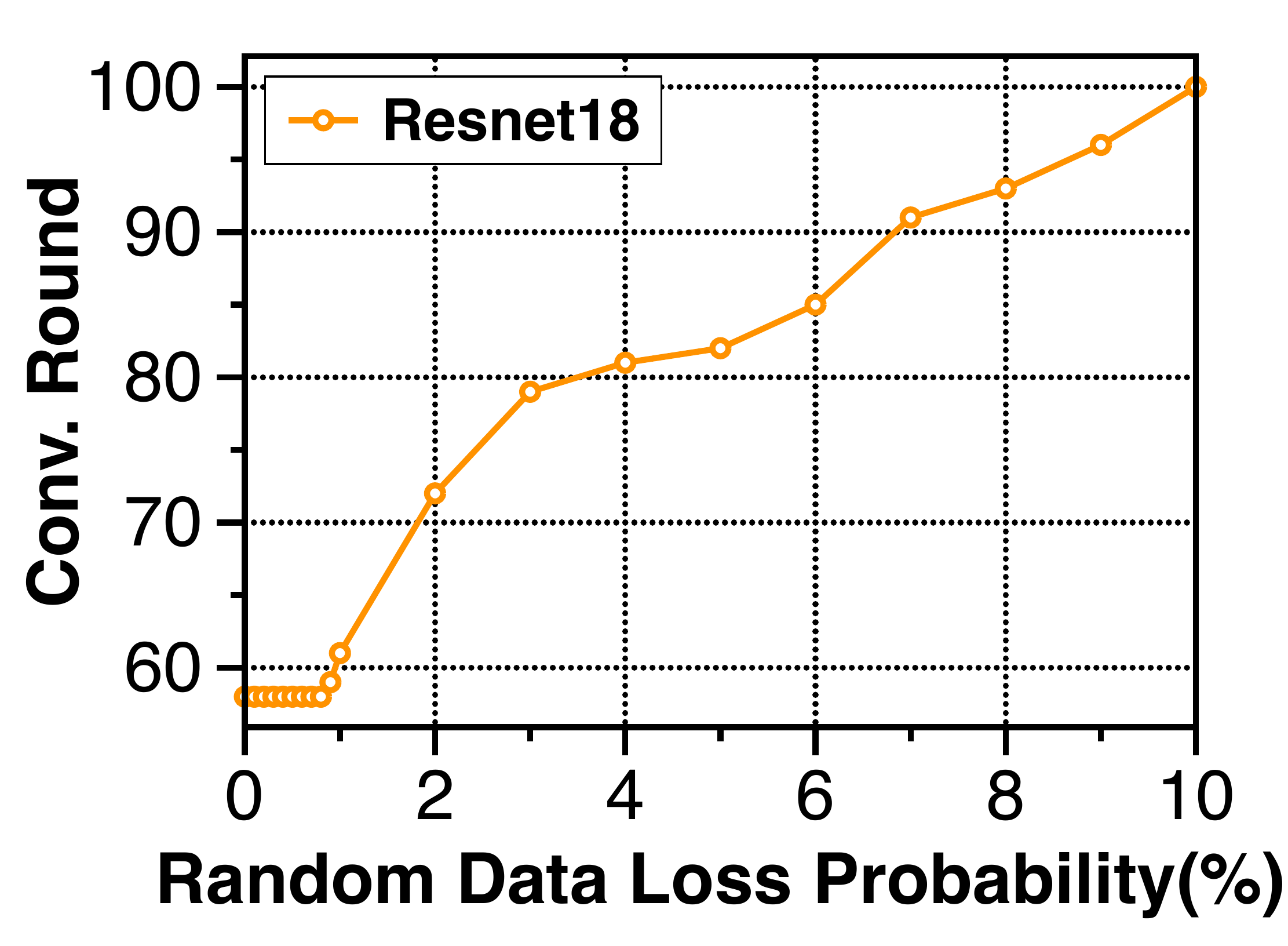}
\end{minipage}
}
\caption{Impact of data loss on model convergence}
\label{fig:model_level}
\end{figure}

\begin{table}[t]
\centering
\small
\begin{tabular}{|p{1cm}|p{0.9cm}|p{1.2cm}|p{1.3cm}|p{1.62cm}|}
\hline
\multicolumn{1}{|c}{Model}   & \multicolumn{2}{|c}{ResNet18} & \multicolumn{2}{|c|}{GRU} \\ 
\hline
\hline
\multicolumn{1}{|c|}{Dataset} & Cifar100  & Caltech101    &  wikitext-2   & wikitext-103\\ \hline
\multicolumn{1}{|c|}{Loss bound} & \multicolumn{1}{c|}{0.9\%} & \multicolumn{1}{c|}{0.9\%}         &   \multicolumn{1}{c|}{1.2\%}            & \multicolumn{1}{c|}{1.4\%}\\ \hline
\end{tabular}
\caption{Loss tolerance bounds across different datasets}
\label{table:different datasets}
\vspace{-4mm}
\end{table}

\parab{Observation 1: DNN training is bounded loss tolerant.} 
Nowadays, DNN training with SGD is essentially an \textit{approximation} algorithm which estimates better parameter values based on information acquired from mini-batches. Such SGD-based training algorithms are \textit{error-tolerant} for two reasons: (1) in each iteration, certain error in parameter-gradient values does not necessarily affect the model accuracy too much, and (2) even an error occurred in earlier iterations can also be sewn up and fixed in later iterations as later iterations start with earlier results. As a result, the error-tolerance feature suggests that certain data loss in communication may not affect model performance! 



To validate the above hypothesis, we inspect the SGD training on different neural network architectures and evaluate the impact of data loss ratio on model convergence. We randomly drop some packets without retransmission and measure the convergence rounds toward the same prediction accuracy. Figure~\ref{fig:model_level} illustrates the example results on the RNN and CNN models. We notice that, for both models, when the data loss ratio is up to $1\%$, the model can converge with the same epochs. Beyond this threshold, when there are more data loss, the convergence speed degrades gradually which means it requires more epochs to converge to the same prediction accuracy. 

We refer to this phenomena as {\em bounded-loss tolerance}: DNN training tolerates a certain fraction $p$ of data loss without affecting the iterations needed for the same accuracy. We further validate such bounded-loss tolerance property across a wide range of DNN models using several general training datasets. Table~\ref{table:bounded_loss} summarizes the bounded loss tolerance ratios for these models, which confirms our hypothesis.  

Furthermore, we note that while different models have different loss tolerance bounds, the bound for the same model across several general training datasets we used remains similar. We show an example in Table~\ref{table:different datasets}, in which the bounds for ResNet18 over Cifar100 and Caltech101 are the same, while the bounds for GRU over wikitext-2 and wikitext-103\cite{caltech101} only differs by 0.2\%. This enables us to profile the loss tolerance bound values for general DNN model architectures\footnote{We note that in practice the bound of a model may vary if datasets differ greatly in some aspects. To explore the loss tolerance bound of a model on a large dataset, one practical way is to use the tolerance bound derived from a smaller sampled sub-dataset from the original dataset as an approximation. Through experiments, we found that the loss tolerance bounds remain almost the same between the sampled sub-dataset and the original dataset.}. 

\begin{figure}[h!]
\subfigure[Gradient loss with different layers]{
\begin{minipage}[b]{0.225\textwidth}
\includegraphics[width=\textwidth]{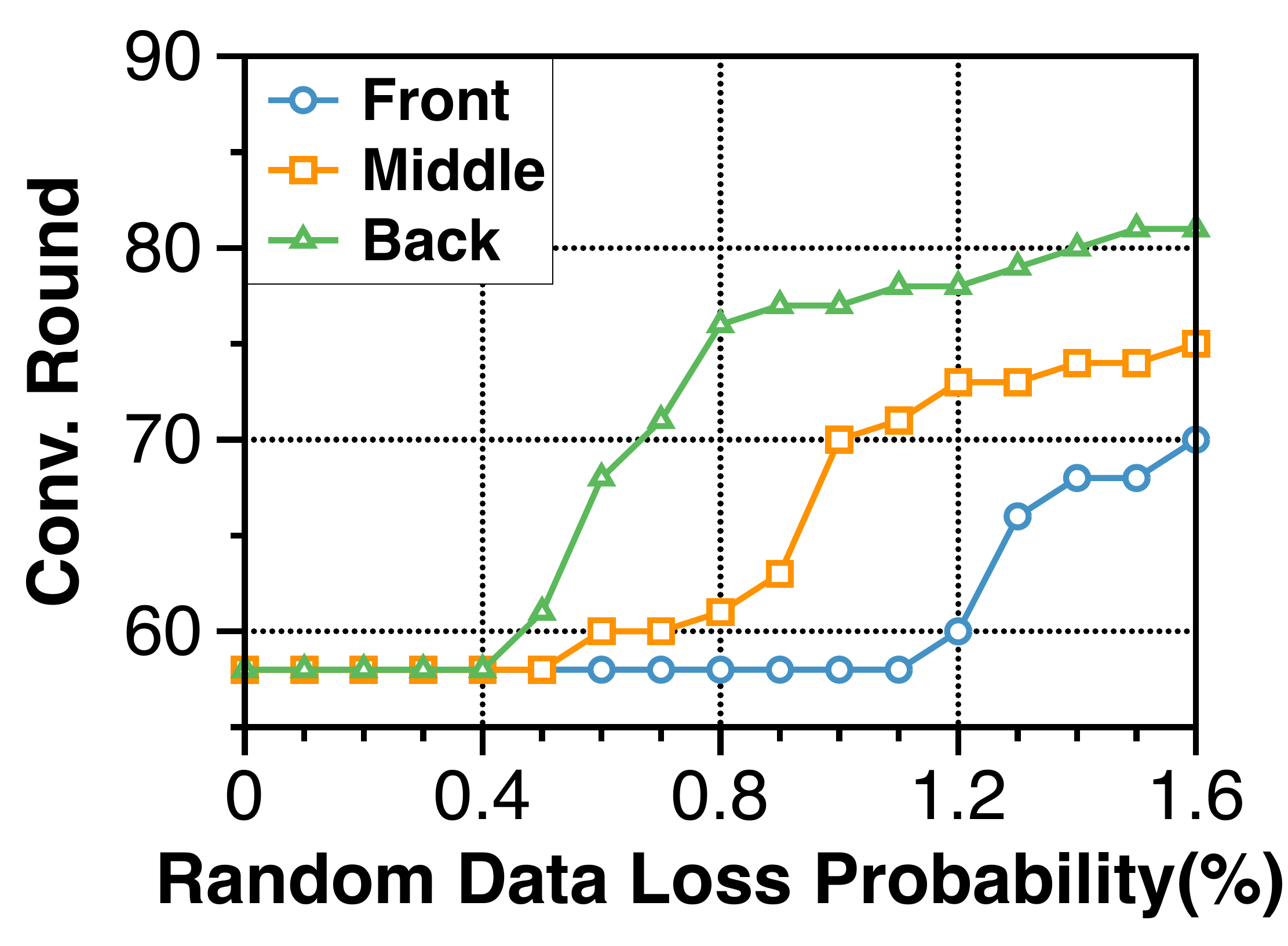}
\end{minipage}
}
\subfigure[Gradient loss with different values]{
\begin{minipage}[b]{0.225\textwidth}
\includegraphics[width=\textwidth]{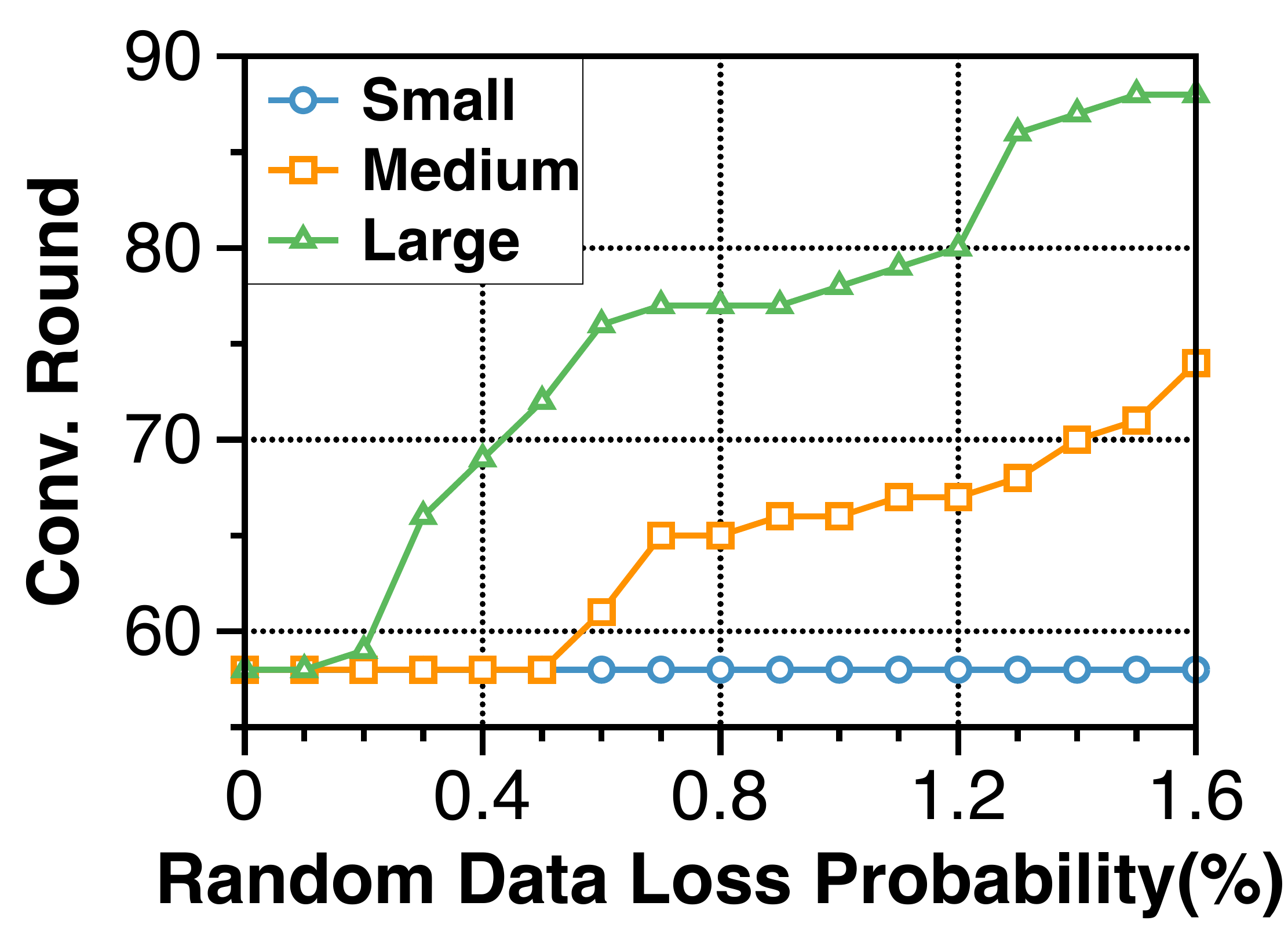}
\end{minipage}
}
\caption{Gradient/packet losses on different NN layers (a) and with different granularity values (b) have different impact to model convergence. }
\label{fig:layer & packet level}
\end{figure}

\parab{Observation 2: Different gradients have different impacts.} Not all gradients are equal. Our insight is that gradients can be differentiated in the following two ways. 

First of all, gradients of front layers are more loss tolerant than back layers in terms of model convergence. In other words, dropping gradients of front layers has less impact than dropping that of back layers. This is because, in deep neural networks, different layers extract features in different levels of abstraction. Generally, back layers contain accumulated information that is learned based upon information in front layers, so is of higher importance~\cite{yosinski2014transferable, lecun2015deep}. For example, front-layer generates low-level general representations like egdes and corners of simple concepts easy to learn, but high-level layers take more steps to learn complex and specific concepts like certain object shapes built upon those simple concepts. In the meanwhile, gradients of front layers are less delay tolerant in terms of training pipelining. This has been observed in prior work~\cite{p3,bytescheduler}. The reason is that the forward propagation (FP) can begin as soon as the front-layer are received, so they should be transmitted earlier, if possible, in order not to delay the pipelining.

To illustrate the impacts of gradient loss on different layers, we conducted an experiment of training ResNet18 in which we randomly discard gradients from: 1) the front layers (the first 20\% layers), 2) the middle layers (the middle 20\% layers), and 3) the back layers (the last 20\% layers) with different loss probabilities. It evident from Figure~\ref{fig:layer & packet level}(a) that gradients in front layers are more tolerant to loss than of the back layers. For example, to maintain the same convergence speed, we can tolerate 1.1\% gradient loss in the front layers but only 0.4\% from the back layers. 

%

Second, gradients of larger magnitude have more impact. This is because, during the training process, SGD leverages gradients to learn the correlations between the intermediate features and the model output. For a given data sample, larger gradients possess stronger correlations between the connected features and the task than of small gradients, and thus are more important for training. As a result, their losses may negatively affect the convergence speed and model accuracy~\cite{lecun2015deep}. In addition, larger gradients indicate bigger learning step size, therefore have more impact on convergence speed.

To show the impact of dropping gradients of different magnitudes, again, we consider three scenarios: randomly dropping gradients among (1) the smallest $20\%$ magnitude, (2) the medium $20\%$ magnitude, and (3) the largest $20\%$ magnitude with different loss probabilities. As shown in Fig~\ref{fig:layer & packet level}(b), it is very clear that dropping larger gradients has more impact than dropping smaller gradients. For example, to maintain the same convergence speed, we can tolerate more than 1.6\% loss of small gradients but only 0.1\% of the large gradients.

\parab{Observation 3: Packets in DNN training are order-independent.} 
Unlike many other classical applications, packets in DNN training are resilient to packet reordering. This is because in classical applications, a message usually consists of multiple packets, and therefore, ordering needs to be maintained among packets of a message. In contrast, in DNN training, multiple messages are packed within one packet, as gradients or parameters are often represented as floating point numbers of 32-bit or less, and a typical packet of 1400-byte payload contains hundreds of such messages. As a result, packets can be interpreted independently and ordering is unnecessary. Such inter-packet order-independency provides opportunity for packet-level load balancing in the network.

\section{Design}\label{sec:design}
We first describes the key ideas behind \sys ($\S$\ref{subsec:ideas}), followed by the detailed \sys mechanisms ($\S$\ref{subsec:mechanism}).


\subsection{Key Ideas}\label{subsec:ideas}
Inspired by the three observations in $\S$\ref{subsec:opportunities}, we come up the following three key ideas correspondingly for optimizing communication for DNN training. 


\parab{Key idea 1: Cutting tail latency with bounded loss tolerance.} 
As introduced in $\S$\ref{subsec:problem}, while solutions such as gradient sparsification\cite{sparse2018nips} or quantization\cite{qsgd2017nips} reduce traffic volume, they do not make communication completely immune to long-tail latencies caused by transient packet drops or queueing. The key reason is that the tail latency is usually caused by the instantaneous traffic pattern such as incast, not only the traffic volume. Even 80\% traffic reduction does not wipe out such long tail (Figure~\ref{fig:motiv:reduce_size_tail}).

To address the problem, we exploit \textbf{observation 1}. Currently, reliability in transport control is ``all-or-nothing'': TCP requires all packets to be received and can thus be blocked by a tiny fraction of packet drops which cause retransmission timeouts; whereas UDP has no reliability guarantee. Neither suits for communication for DNN training. Based on observation 1, we propose a simple yet effective bounded-loss tolerant end-host transport protocol that minimizes the data transmission time, by intentionally ignoring packets (bounded by $p$) delayed or lost in the network without retransmissions. This effectively cuts the tail latency by avoiding costly retransmission timeouts. 



\parab{Key idea 2: Optimizing training efficiency with gradient-aware queueing/dropping.} 
While DNN training process tolerates certain packet losses, the influence of losing different gradients may differ remarkably as shown in \textbf{observation 2}:
\begin{icompact}
\item In terms of layer of gradient, front layer gradient is more tolerant to loss than back layer gradient, whose dropping has less impact on model convergence. 
\item In terms of magnitude of gradient, large gradient is less tolerant to loss than small gradient, whose dropping has more impact on model convergence. 
\end{icompact}
To take advantage of observation 2, when the switch queue is full and some packets have to be dropped, instead of random dropping, we propose a gradient-aware selective dropping:
\begin{icompact}
\item Packet carries front layer gradients will be prioritized for dropping than that carries back layer gradients.  
\item Packet carries larger gradients will be de-prioritized for dropping than that carries smaller gradients. 
\end{icompact}

Furthermore, as pointed out in $\S$\ref{subsec:opportunities}, while gradients of front layers are more loss-tolerant, they are less delay tolerant in training pipelining, since forward propagation (FP) can begin as soon as the front-layer tensors are received. Therefore, in addition to selective dropping, we further enforce priority queueing to prioritize front-layer packets. Later, we show that both selective dropping and priority queueing can be implemented together with commodity switches ($\S$\ref{sec:impl}). In Appendix~\ref{sec:analysis:converge}, we give a mathematical proof to show that by this mechanism, convergence is guaranteed even with gradients loss.

\parab{Key idea 3: Enabling per-packet load balancing with inter-packet order-independency.} 
Load balancing tries to eliminate hotspots by spreading traffic on multiple paths. Ideally, this should be done at packet-level. However, current practice still remains at flow-level (or at most flowlet-level) with sub-optimal performance~\cite{pfabric2013sigcomm,pias,conga,hermes}. One key concern is due to the reordering problem.
Based on \textbf{observation 3}, packets of DNN training are free of ordering among each other, which enables packet-level spreading without reordering issues. Therefore, we propose per-packet load balancing in \sys to fully utilize bandwidth in the network.

\begin{figure}[htbp]
    \centering
    \includegraphics[width=1.05\linewidth]{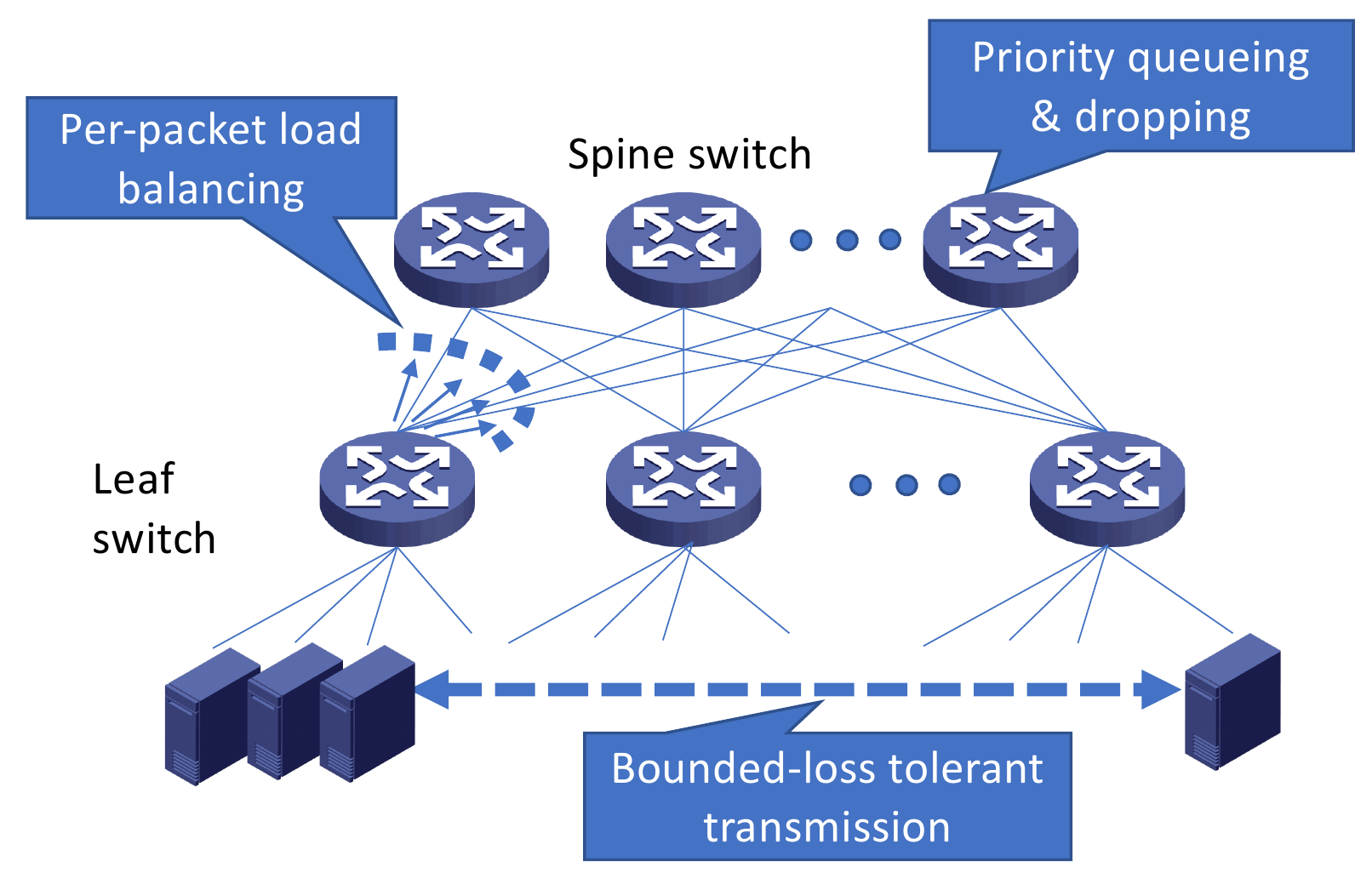}
    \caption{\sys Overview}
    \label{fig:design:overview}
\end{figure}

\subsection{Design Details}\label{subsec:mechanism}
We proceed to introduce the design details of \sys which integrate the above three ideas. Figure~\ref{fig:design:overview} presents an overview of \sys. The general workflow is as follows. For transmission, data are first spread onto multi-path in the network on a per-packet basis to minimize hotpots (\textbf{idea 3}). Then, if congestion happens, the \sys switch will perform priority queueing and selective dropping based on layers and magnitudes of gradients to optimize training efficiency (\textbf{idea 2}). Finally, a bounded loss-tolerant data transmission protocol is established between end-points to avoid long tail latencies caused by retransmission timeouts (\textbf{idea 1}).

\begin{figure}[t]
    \centering
    \includegraphics[width=0.97\linewidth]{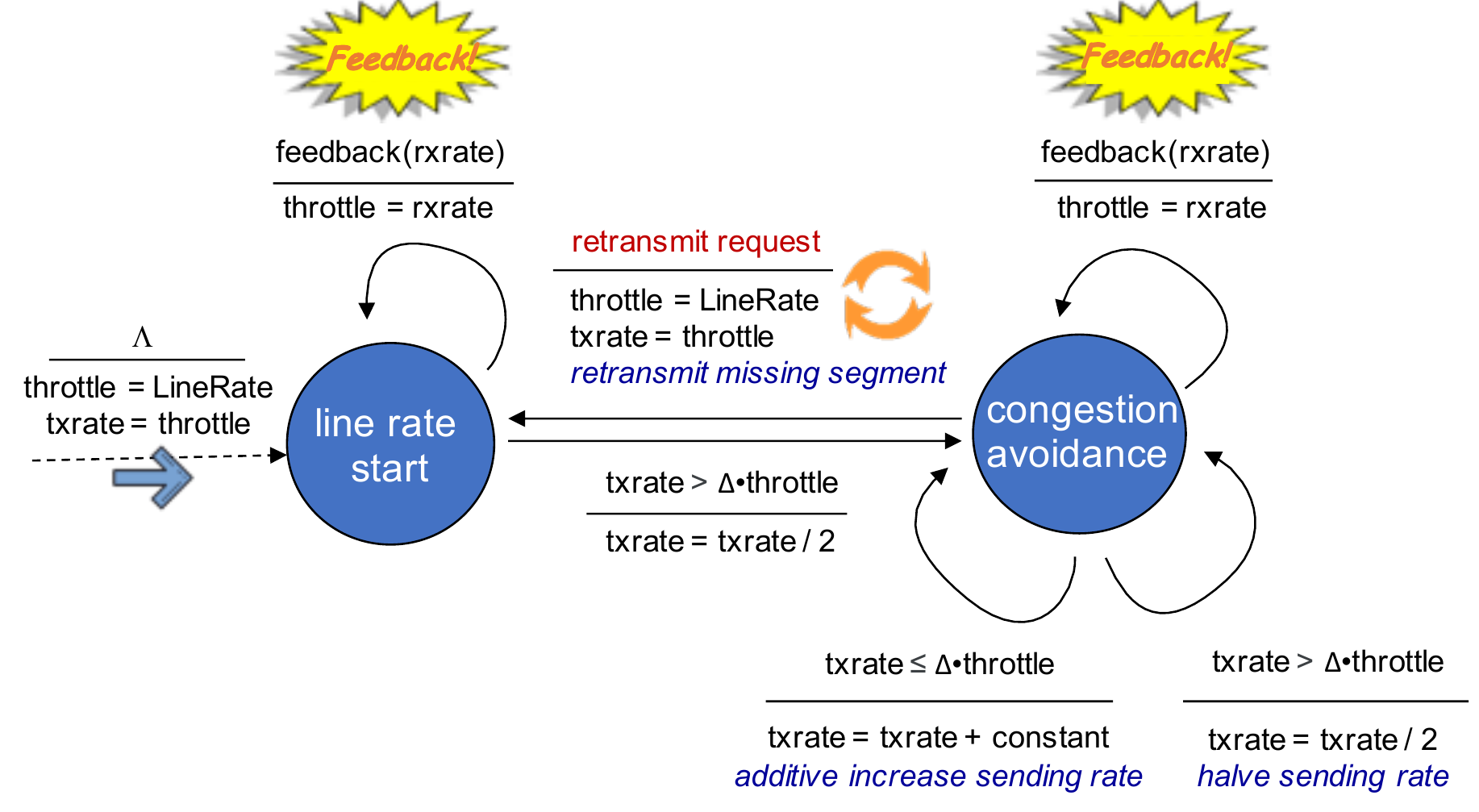}
    \caption{Rate Control State Machine}
    \label{fig:design:statemachine}
\end{figure}

\subsubsection{Bounded-loss Tolerant Data Transmission} \label{sec:design:key1}

We design a bounded-loss tolerant data transmission protocol at application layer, here we describe the mechanisms.

\parab{Zero-RTT connection.} 
To initialize a flow, we first setup a connection, and senders and receivers rendezvous with tensorID, size and loss-bound. To minimize latency, we send gradients at the same time. Notice that this may pose correctness issues since receivers may not prepare enough buffer or initialize the state. Fortunately, DNN communication traffic is fixed and repetitive, we can set the buffer size as the largest tensor before communication starts. Meanwhile, we discard gradients arrival earlier than the rendezvous information.

%

\parab{Loss tolerant transmission.} We transmit gradients in an unreliable channel to achieve high throughput and free of packet loss or out-of-order issues, and transmit signal packets in a reliable channel to guarantee robustness. Meanwhile, we use a higher priority switch queue for signal packets.
During the transmission, \sys sender keeps transmitting  gradients of a tensor until it receives a flow finishing signal or the gradients are sent out. If there are no more gradients to send, the sender emits a flow stop signal immediately. 

\parab{Guarantee loss bound retransmission.} 
To guarantee receiving enough gradients, \sys receiver checks whether the bound requirement is met on receiving data. If the receiver is notified with a flow stop signal, it requests a retransmission for the missing gradients, and the sender will go through loss tolerant transmission again.
After the receiver meets its bound requirement, it sends a flow finishing signal and keeps receiving on-the-fly packets until a flow finishing confirm from the sender reaches.

\parab{Minimal rate control.} In virtue of the loss-tolerance feature, our data transmission protocol needs a minimal rate control only to achieve high throughput without congestion collapse. Figure~\ref{fig:design:statemachine} uses a state machine to illustrate the procedure:
\begin{icompact}
	\item Initially, flow sends at the line rate, which is equal to link bandwidth. Receivers periodically send measured packet receiving rate to senders, if the sending rate is larger than the receiving rate times a factor $\Delta$, we change the state to "congestion avoidance" and halve the sending rate.\footnote{In our implementation, the period is 200 $\mu$s, $\Delta$ is equal to 2.} 
	\item During the "congestion avoidance" state, we compare the packet sending rate with the measured packet receiving rate every period. When the former is smaller than $\Delta$ times the latter, we half the sending rate, to avoid congestion collapse. Otherwise, we do additive increase to the sending rate, which is 5\% of the link bandwidth in our setting.
	\item When the flow is sending out and requests to do retransmission, we change the state to "line rate start" and reset the sending rate to link bandwidth.
\end{icompact}

\parab{Loss bound setting.} Our design of the transmission interface allows per-tensor a loss bound. \footnote{For convenience, we apply a global bound for all the tensors in a model and leave the different loss bound setting to future work.} The bound set here is just a guarantee that at least the amount of data will be received. 
Besides, parameter synchronization can be decomposed into a push stage and a pull stage (corresponding to a reduce plus a broadcast in all-reduce communication). We distinguish push from pull and only enforce a lower loss-tolerant bound for pull stage because each gradient in the pull stage is in nature aggregated from many workers, thus more important.

\begin{figure}[t]
    \centering
    \includegraphics[width=0.97\linewidth]{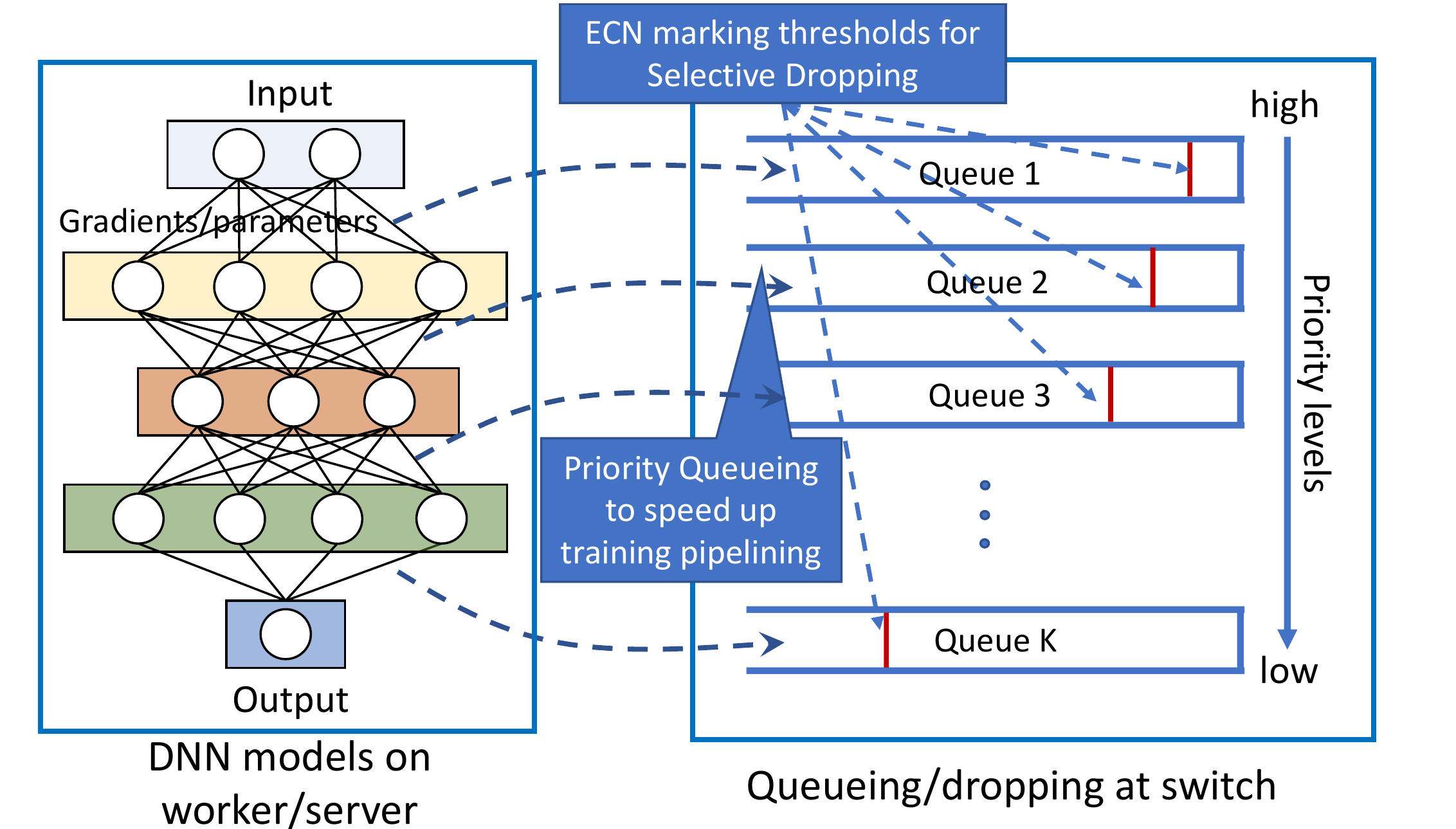}
    \caption{\sys Switch}
    \label{fig:design:f322}
\end{figure}

\subsubsection{Gradient-aware Packet Queueing and Dropping} \label{sec:design:key2}




To enforce this gradient-aware traffic scheduling, \sys first tags packets at end-host with two level of information: layer of gradients and magnitude of gradients. In \sys switch, to enforce the priority queueing, packets are added to different priority queues according to the layer of gradients. To enforce the selective dropping, small/front layers gradients are dropped earlier before the switch buffer is full. Here, we give the design of each part in detail.

\parab{End-host packet tagging.} 
To encode the layer of gradients in a packet, a straightforward idea is to map each layer to a unique priority. However, it is impractical because the number of layers can be much more than switch priorities (typically 8). To address it, we evenly map all the layers to available priorities. Specifically, we tag the packets of $x$-th layer with $\frac{xP}{L}$, where $L$ and $P$ are the total number of layers and priorities respectively. 
To encode the magnitude of gradients information into priorities, 
given multiple gradients in a packet, we calculate the average of the gradient magnitudes and compare it with a threshold to determine whether the packet should be marked as important. By default, we set the threshold to the median value of all gradient's magnitude of each tensor. To reduce the median value calculation time, we sample only 0.1\% of the gradients. 





\parab{\sys switch.} The \sys switch performs priority queueing~\cite{bai2017pias, zhang2019enabling, chen2018auto} and selective dropping for packet scheduling, as shown in Figure~\ref{fig:design:f322}. For priority queueing, the switch first maps packets of front layers to high priorities. Then, it adopts the standard strict priority queueing discipline. This straightforwardly speeds up training pipelining.
For selective dropping, the switch decides whether to drop a packet in a hierarchical manner. On layer level, the switch checks the layer information and pushes the packet to different priority queues. To prioritize the dropping of front layer packets, the corresponding queues are set with lower dropping threshold (We give an analysis of the dropping threshold setting in Appendix~\ref{sec:analysis:thres}). On magnitude level, the switch decides whether to drop the packets based on the importance of the packet. Only unimportant packets will be selectively dropped. To implement the selective  dropping on commodity switch, we use ECN marking thresholds as the dropping threshold, the detail is described in $\S$\ref{sec:impl:switch}.



\begin{figure}[htbp]
    \centering
    \includegraphics[width=1\linewidth]{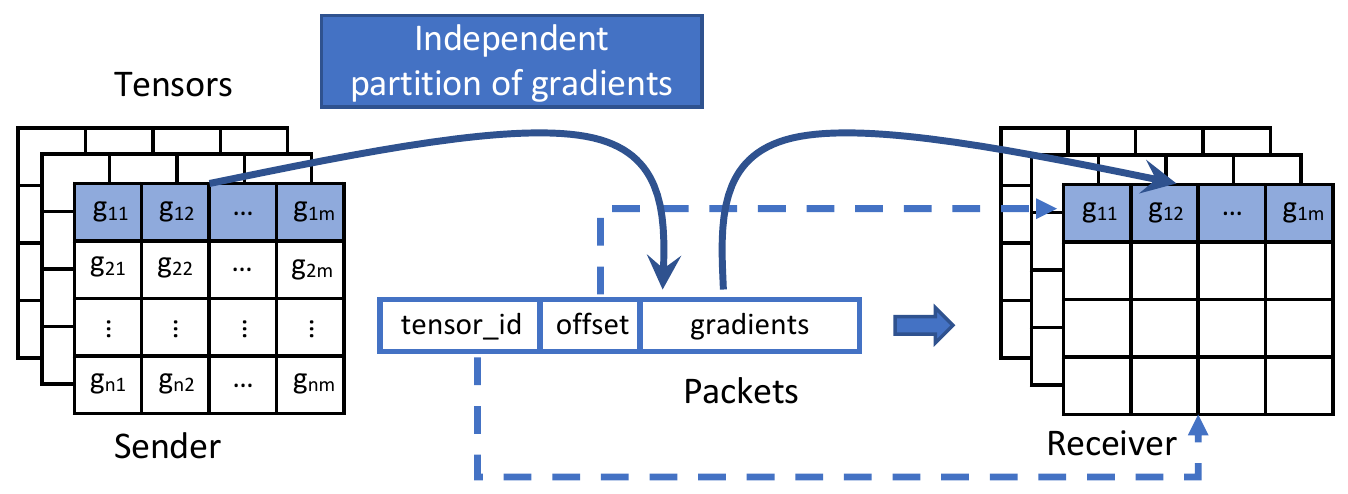}
    \caption{Tensor Partition \& Reconstruction}
    \label{fig:design:f323}
\end{figure}

\subsubsection{Per-packet Load Balancing} \label{sec:design:key3}

\sys performs per-packet load balancing, to fully achieve inter-packet order-independency and accelerate the processing speed of received packets, we carefully do tensor partition and reconstruction.


\parab{Tensor partition \& reconstruction.} Before transmission, each tensor is divided into independent partitions, the "independent" means no gradient is across two packets, the "partition" means a segment of consecutive gradients in the tensor. As we can see from Figure~\ref{fig:design:f323}, each independent partition is packed into one packet (suppose one packet can contain exactly m gradients), with its address information (which tensor, offset in the tensor). When packets arrive at the receiver, it will be immediately placed into the right place according to its address information, without being impacted by the arrival order. Meanwhile, even if some packets are lost, the tensors can be reconstructed with retaining most of the information. 


\parab{Load balancing.} \sys spreads data packets evenly among multiple parallel paths between source and destination. We provide two design choices. One design choice is to leverage the switch side per-packet ECMP\cite{ecmp} or the so-called packet spraying~\cite{dixit2013impact}. This is simple to implement and deploy. It works well for a symmetric network topology, which is typical in datacenter. The other choice is to give the end-host the control of multi-path routing~\cite{hu2016explicit}. This can be done by source routing or label switching.  While it adds a bit overhead at the end-host to collect the routing information, the strength is that it works well even for an asymmetric topology (or symmetric topology with link failures). 

%

%
%

\section{Implementation}
\label{sec:impl}

\begin{figure}
    \centering
    \includegraphics[width=1.0\linewidth]{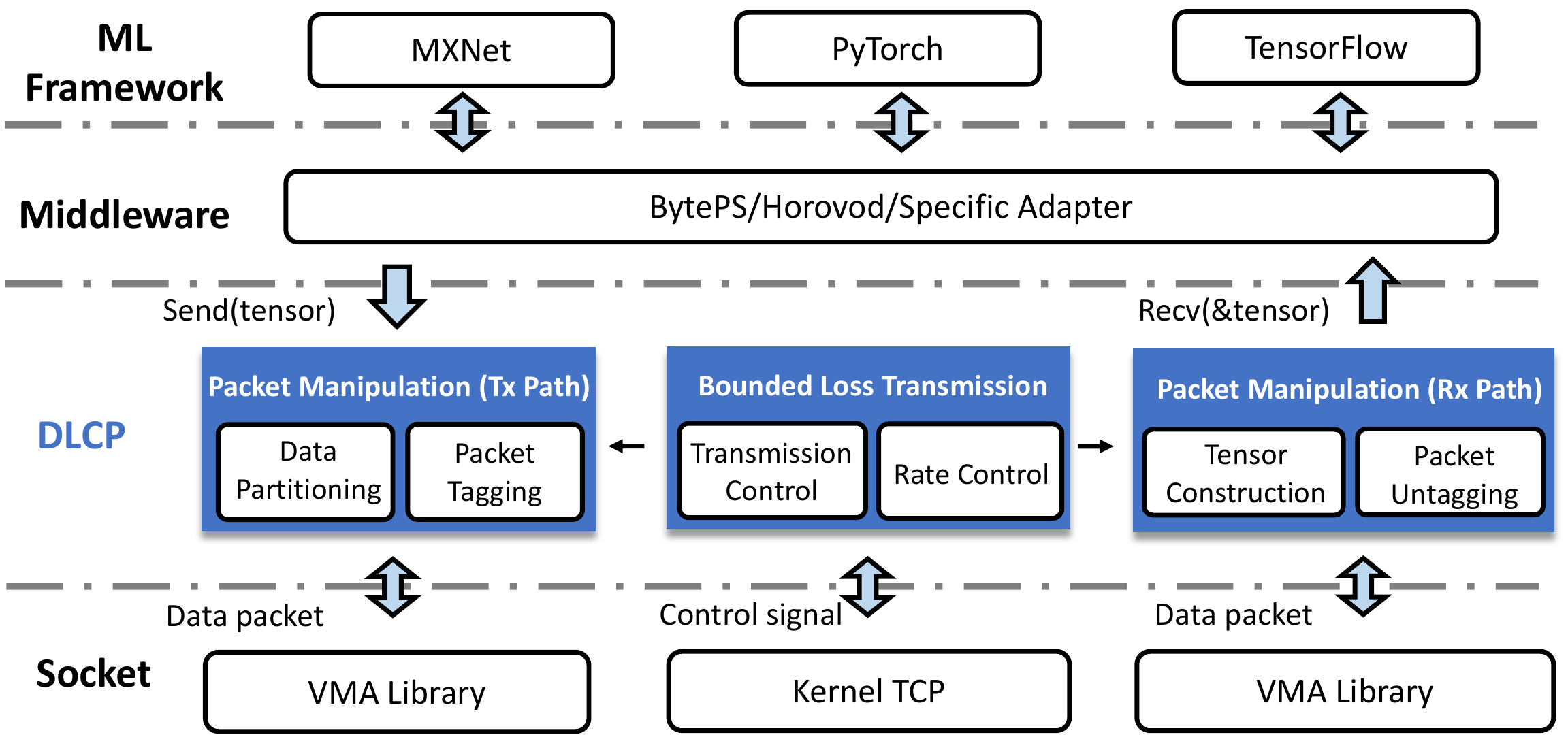}
    \caption{\sys End-host Implementation Overview}
    \label{fig:impl:header}
\end{figure}

We build a prototype of \sys using Mellanox LibVMA~\cite{libvma} and commodity switches and integrate it to popular ML frameworks likes Tensorflow\cite{tensorflow}, Pytorch\cite{pytorch} and MXNet\cite{mxnet}.

Here we describe the details of each component.
%
\subsection{End-host Network Stack}

\parab{Overview.} 
As shown in Figure~\ref{fig:impl:header}, \sys is implemented between machine learning framework layer and socket layer.
We provide a series of universal communication interfaces to enable bounded loss tolerance and packet tagging. The interfaces are flexible and can be integrated into various ML frameworks~\cite{tensorflow,pytorch,mxnet}
and distributed training middleware systems (e.g., Horovod \cite{horovod} and BytePS \cite{bytescheduler}) without needing to modify their operating system kernel. To this end, we design and implement basic communication primitives in user space based on common application abstractions of these ML frameworks.
Our prototype demonstrates that \sys can support more datapaths in the future such as user-space network stack, DPDK, RDMA UD \cite{rdmaud}, Cisco usNIC \cite{usnic}, and hardware datapaths.


\parab{Universal interfaces.} We provide two basic communication primitives \texttt{dlcp\_send(tensor, prio\_func)} and \texttt{dlcp\_recv(tensor, loss\_bound)}. A tensor is essentially a memory space that stores gradients and some metadata (e.g., shape and data type). The tensor abstraction has been widely used by almost all popular DNN frameworks, e.g., Tensor in TensorFlow and PyTorch, and NDArray in MXNet. 



\parab{\sys sender.} On the sender side, a tensor is first partitioned into some consecutive MTU-sized (excluding header overhead) segments of gradients. Then we run the priority function \texttt{prio\_func} for each gradient segment to calculate its priority. Finally, we add an Ethernet header, an IP header, a UDP header, and a \sys header to each segment to form a UDP packet. A \sys header encodes the tensor identifier, length, offset and a sequence number. Priority is mapped to DSCP value encoded in IP header.


\parab{\sys receiver.} On the receiver side, \texttt{recv} takes as input which tensor to receive and loss-tolerant bound of the tensor. Before data transmission, sender and receiver do a rendezvous to allocate receiving buffer in advance. On receiving a new packet, the receiver copies its gradients to pre-allocated memory buffer according to its offset. The receiver uses a bitmap to maintain the already received gradients.



\parab{Data \& signal transmission.} We implement \sys network stack using both UDP and TCP. Inspired by~\cite{rbudp}, we separate data transfers and control signals, and only provide full reliability for control signals whose traffic size is much smaller. The control signals include flow start/finish, retransmission request and stop request/confirm. To ensure reliability, we use TCP in Linux kernel to carry control signals. To minimize the losses of the control signals, we reserve a separate priority for them at the switch. We find that control packets are rarely dropped in practice. We implement data transfer mechanism using UDP. To achieve high throughput, we adopt UDP in Mellanox LibVMA \cite{libvma} (instead of Linux kernel), a high performance user space network stack. Since our implementation only requires unreliable messaging, \sys can also have other datapaths such as RDMA UD \cite{rdmaud} and Cisco usNIC~\cite{usnic} transports which are essentially OS-bypass low latency UD.

\subsection{Switch Configuration}\label{sec:impl:switch}
We implement the priority queueing and dropping using built-in functions of commodity switches.


\parab{Priority queueing.} We classify packets based on the DSCP field~\cite{pias, li2017rate, hu2017tagger, chen2016scheduling} and map them to the corresponding switch priority queues. We enable strict priority queueing to schedule packet transmissions at the egress.

\parab{Selective dropping.} Current switching chips cannot \emph{push out} packets that are already stored in the switch buffers. Therefore, to realize selective dropping~\cite{hu2020aeolus}, we can only selectively drop packets at the ingress.
To this end, we use RED/ECN function ~\cite{bai2016enabling, zeng2017combining}, which is widely supported by commodity switches~\cite{mellanox}. In current switch implementations, when the switch queue size exceeds the ECN marking threshold, the switch will mark the arrival ECN-capable packets and \emph{drop} not ECN-capable packets. Hence, at the sender side, we only tag the packets carrying significant gradients with ECN-capable.
To implement layer-wise priority dropping, i.e. packets from front layer are easier to drop than packets from back layer, we set lower threshold on higher priority queues. More specifically, the thresholds of the queues are set to an arithmetic sequence $T * (1, 1 + d, ... , 1 + 6d)$. The highest priority queue is reserved for control signals in practice.

\subsection{ML framework integration.} 

\sys can be directly integrated with deep learning frameworks such as TensorFlow~\cite{tensorflow}, PyTorch~\cite{pytorch}, and MXNet~\cite{mxnet} or indirectly integrated with some distributed training middleware systems such as Horovod \cite{horovod} and ByteScheduler \cite{bytescheduler}.
Modern deep learning frameworks have their own distributed training implementation. They tend to choose specific RPC or messaging library and build an abstraction over it. For example, MXNet uses PS-Lite~\cite{ps2014osdi} and build a key-value store over it; PyTorch prefers collective communication API and can have multiple backends such as Gloo~\cite{gloo}, MPI\cite{mpi} or NCCL\cite{nccl}; TensorFlow is more monolithic to support both parameter server and distribute strategy as its communication abstraction and underlay is gRPC. These communication abstraction layers decide which nodes are communicating with each other in one iteration and is usually built on top of point-to-point communication primitives. They provide the flexibility for being implemented with different RPC or messaging libraries. 
To direct integrate \sys with some specific frameworks such as MXNet which use PS-Lite, we only need to implement its abstraction for point-to-point communication with \sys's interface without changing  the existing user code.



%

\section{Evaluation}\label{sec:eval}

We evaluate \sys by using a combination of testbed experiments and large-scale simulations. Our key findings are summarized as follows:
\begin{icompact}
\item In testbed experiments, \sys can accelerate training of a state-of-the-art DNN scheduler by up to 84.3\% without adverse effect on model convergence or accuracy.
\item \sys achieves 11-84.3\% speedup on a range of DNN models, ML frameworks (TensorFlow\cite{tensorflow}, MXNet\cite{mxnet}, PyTorch\cite{pytorch}) and  synchronization schemes (PS\cite{ps2014osdi} vs. AllReduce\cite{allreduce}).
\item \sys significantly reduce the tail FCT. In large-scale simulations, \sys reduces tail FCT up to 91.8\% in a 144-node, 100G network.
\end{icompact}


\begin{figure*}[h!]
    \captionsetup{skip=0pt}
    \centering
    \subfigure[ResNet50]{
	\includegraphics[width=0.23\linewidth]{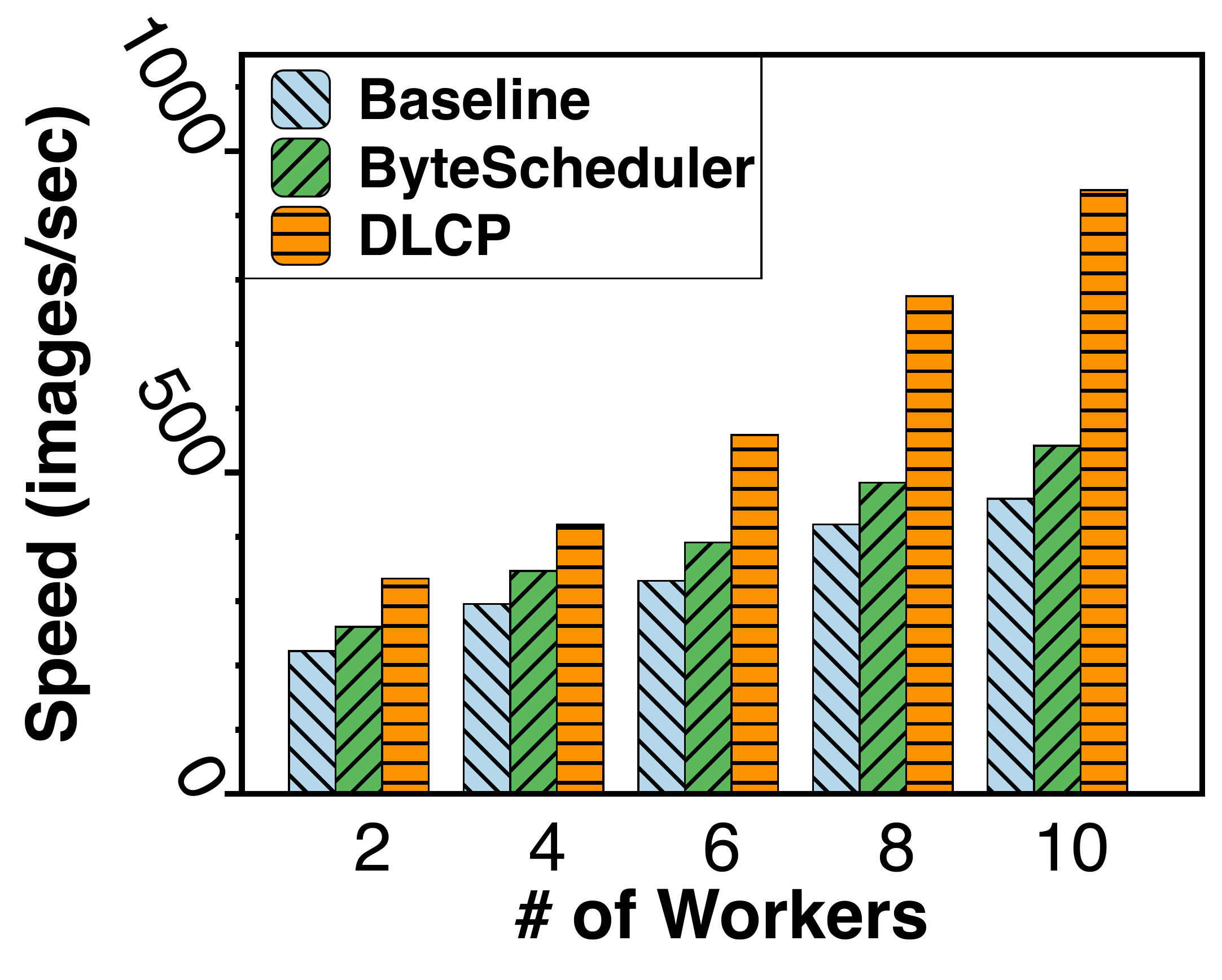}
    }
    \subfigure[VGG16]{
        \includegraphics[width=0.23\linewidth]{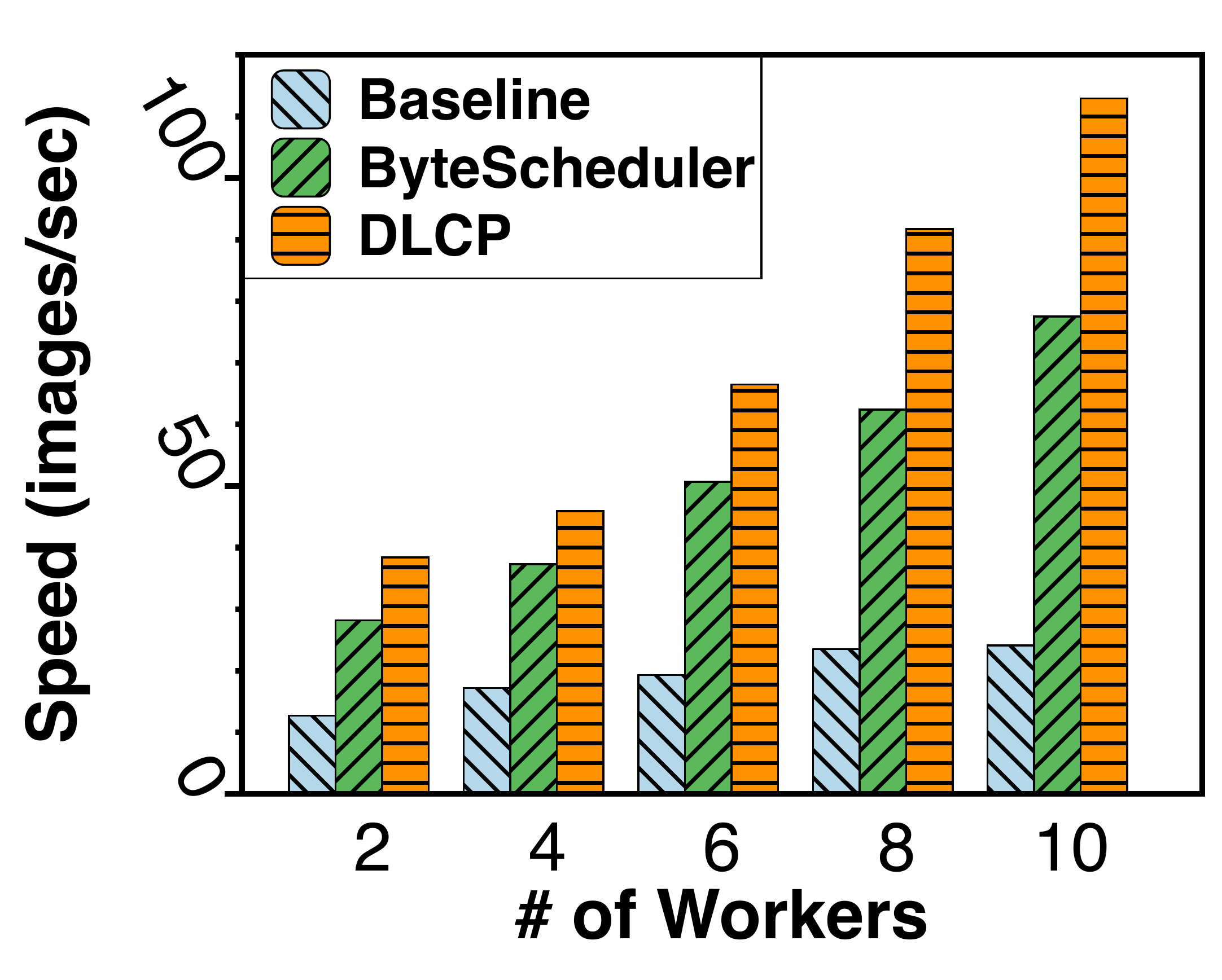}
    }
    \subfigure[Inception-v3]{
	\includegraphics[width=0.23\linewidth]{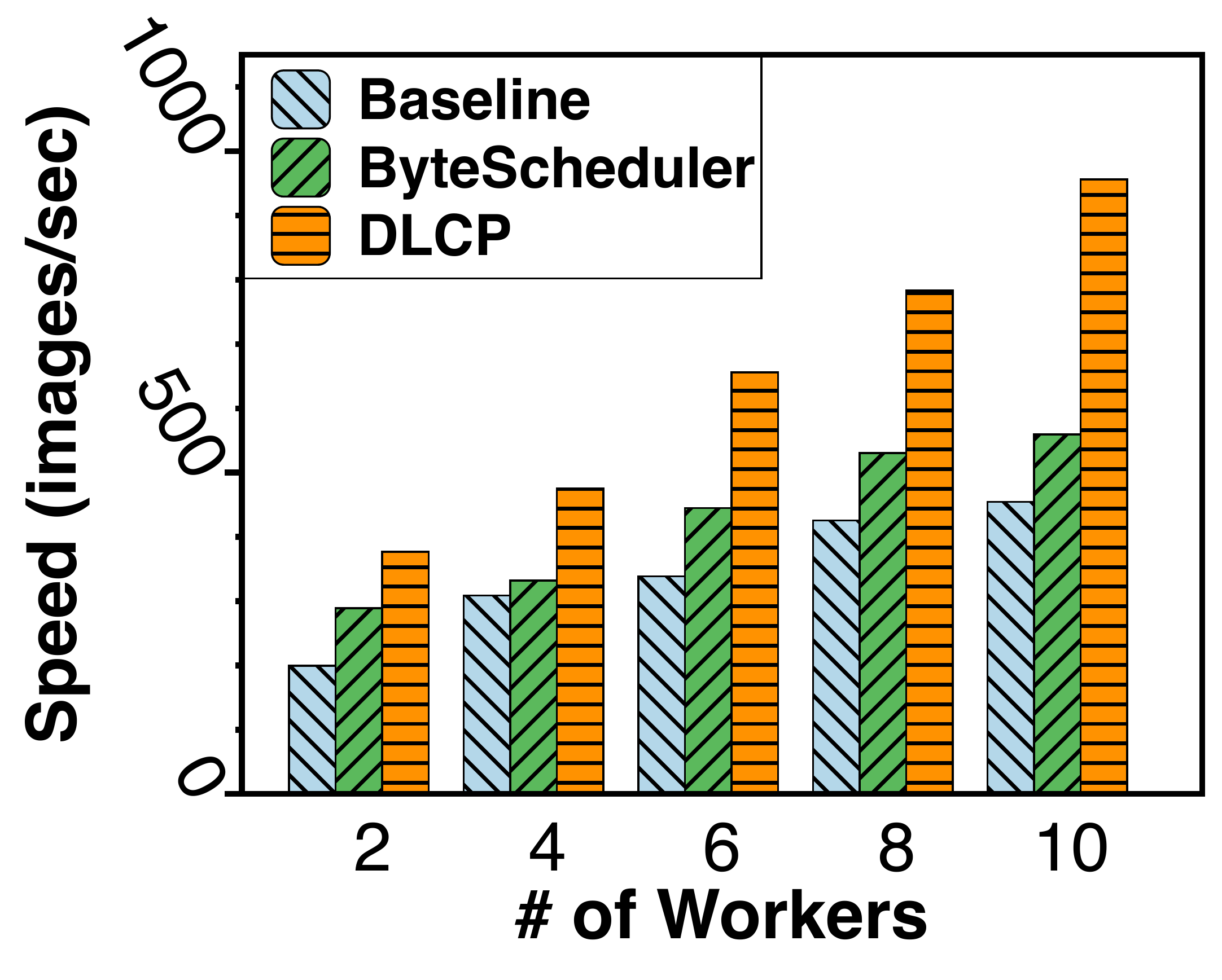}
    }
    \subfigure[Transformer]{
	\includegraphics[width=0.23\linewidth]{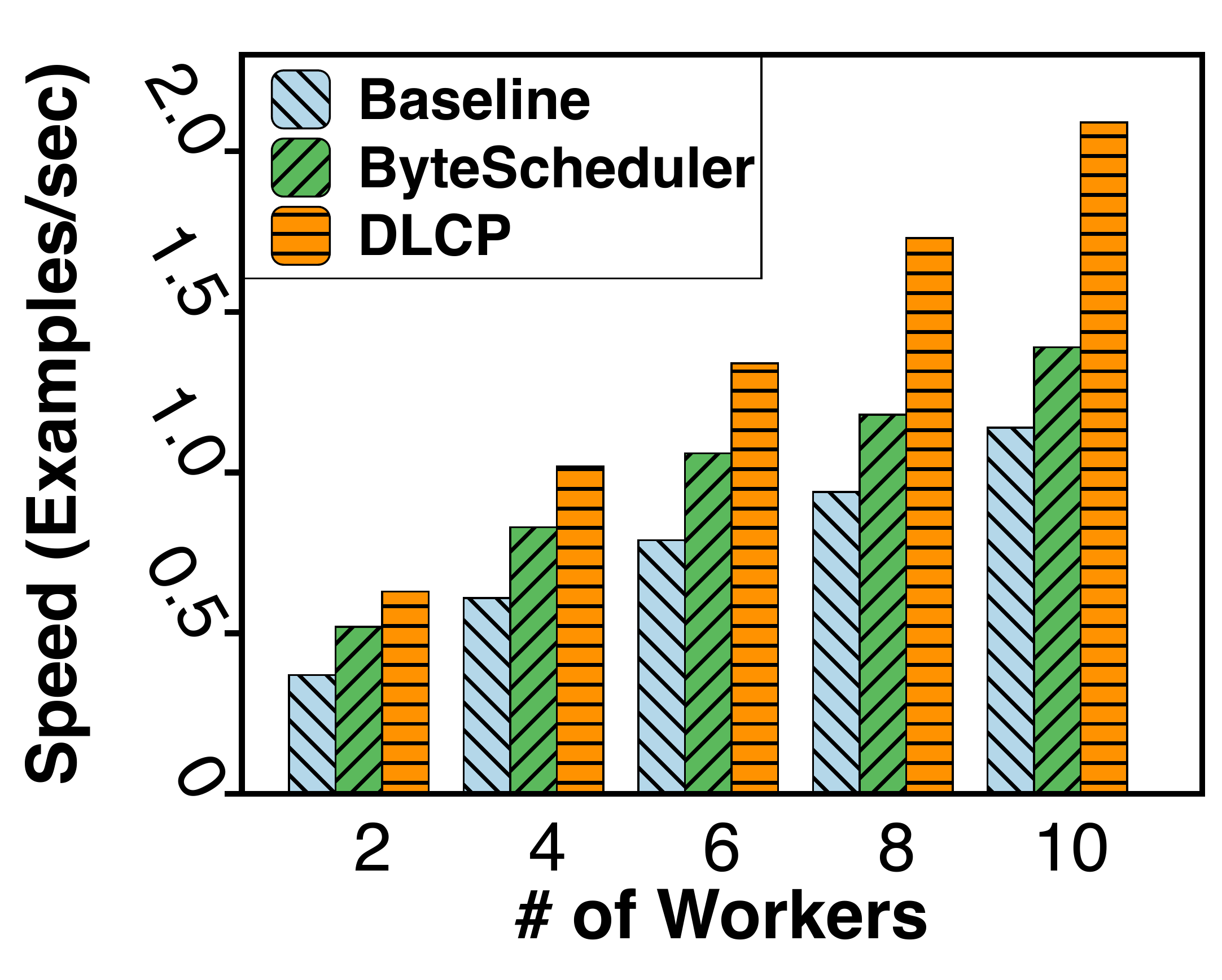}
    }
    \caption{Speedup with Different Models.}
    \label{fig:testbed:models}
\end{figure*}

\begin{figure*}[htbp!]
    \centering
    \subfigure[MXNet, ResNet50]{
        \includegraphics[width=0.23\linewidth]{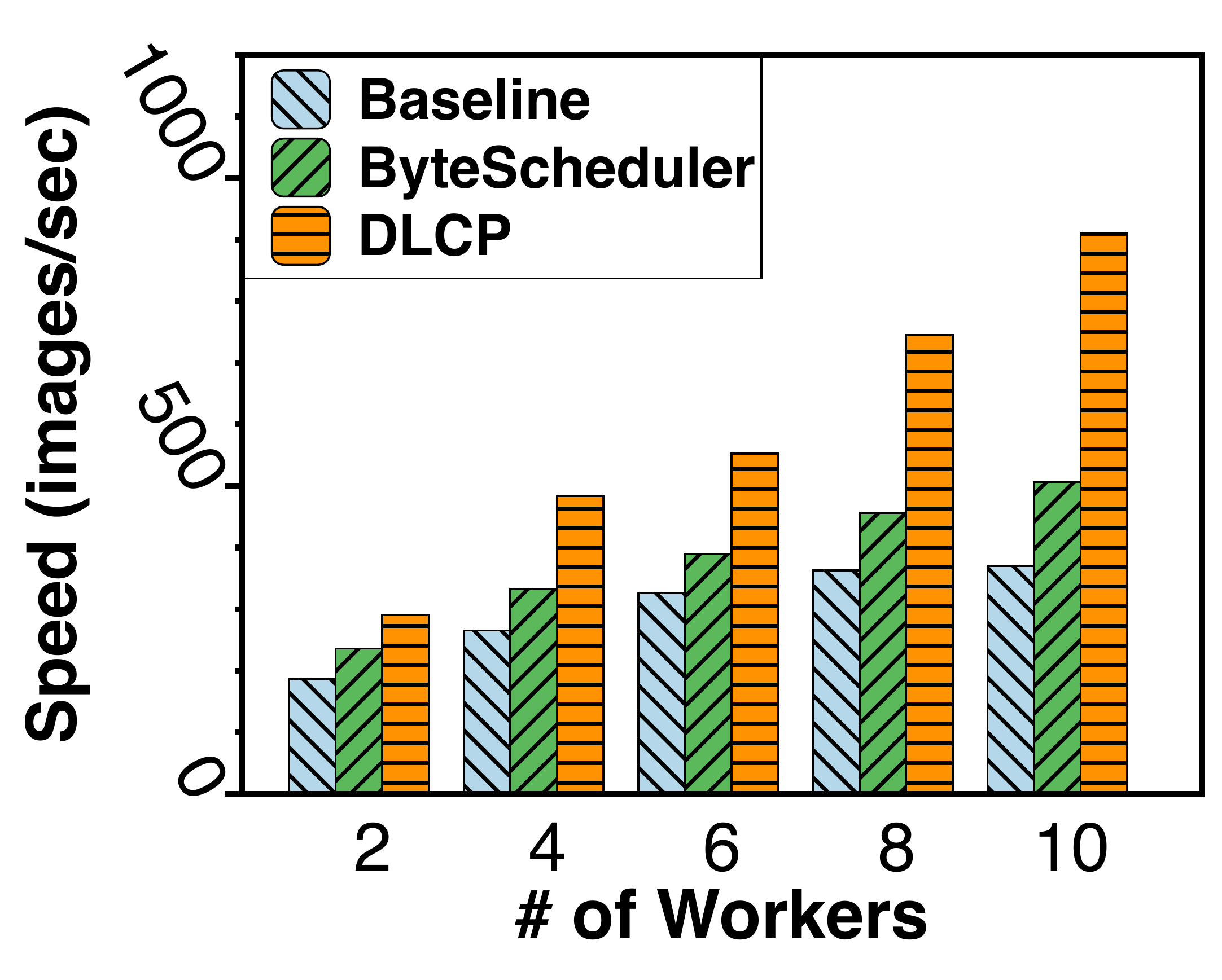}
    }
    \subfigure[PyTorch, ResNet50]{
	\includegraphics[width=0.23\linewidth]{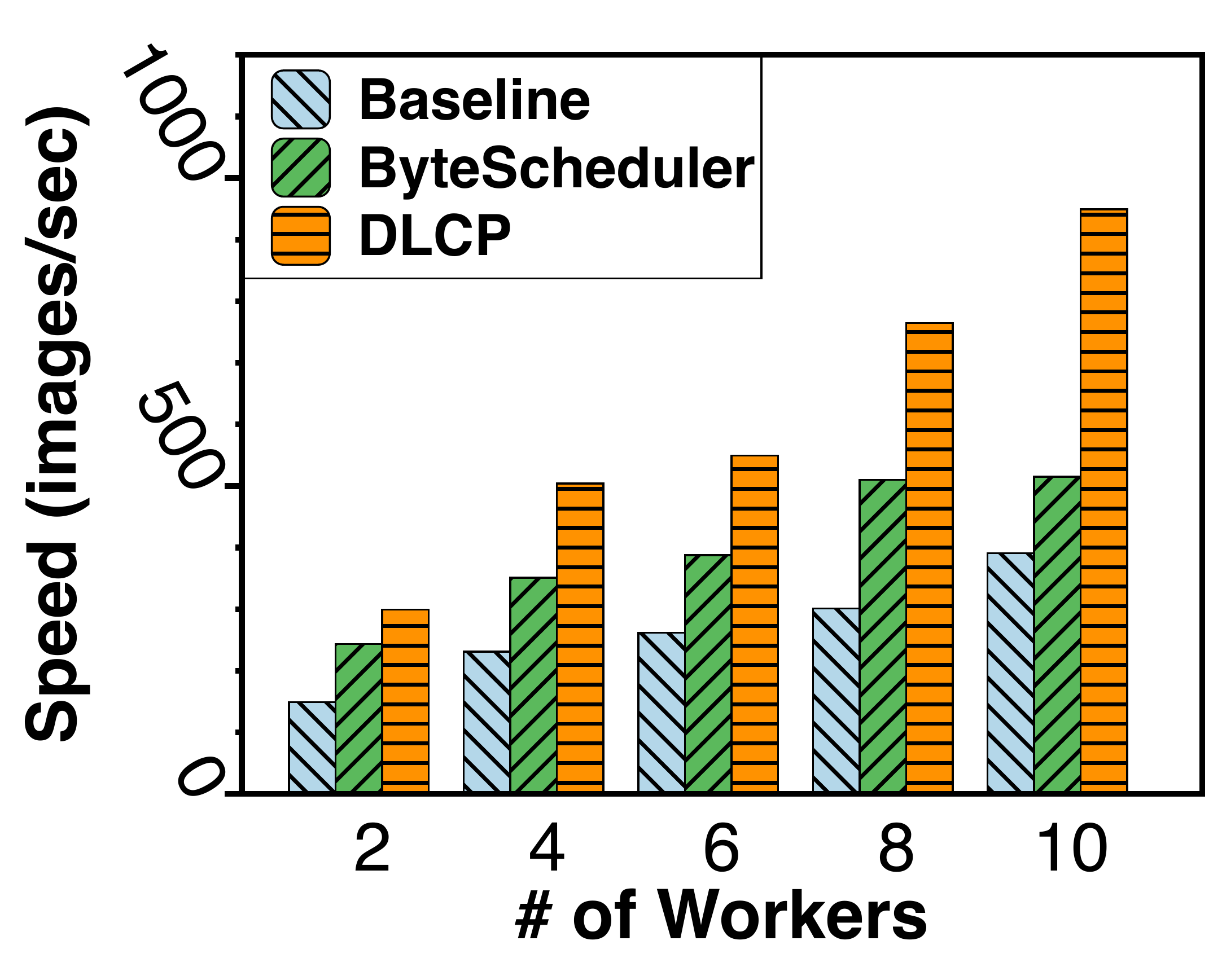}
    }
    \subfigure[MXNet, VGG16]{
	\includegraphics[width=0.23\linewidth]{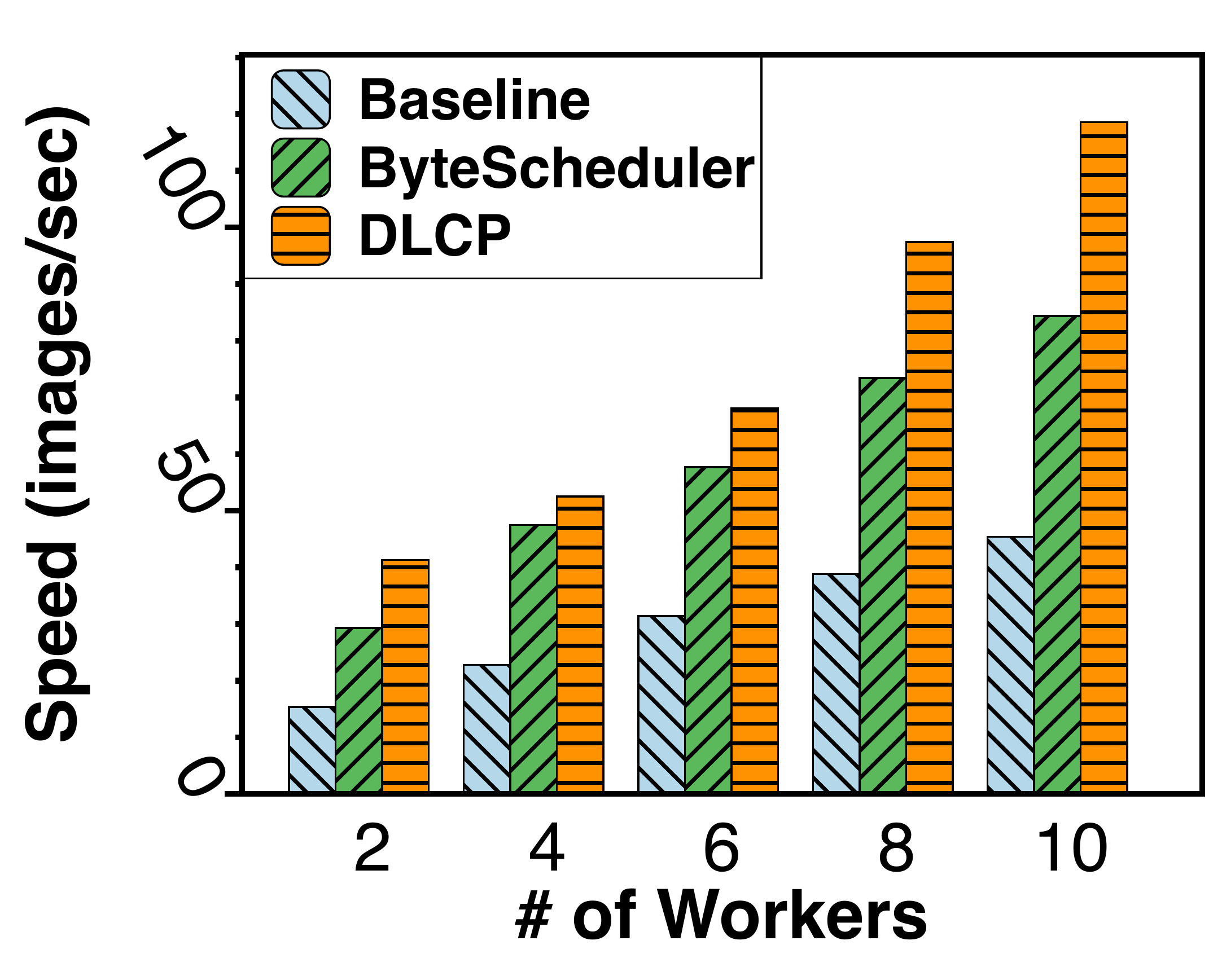}
    }
    \subfigure[PyTorch, VGG16]{
	\includegraphics[width=0.23\linewidth]{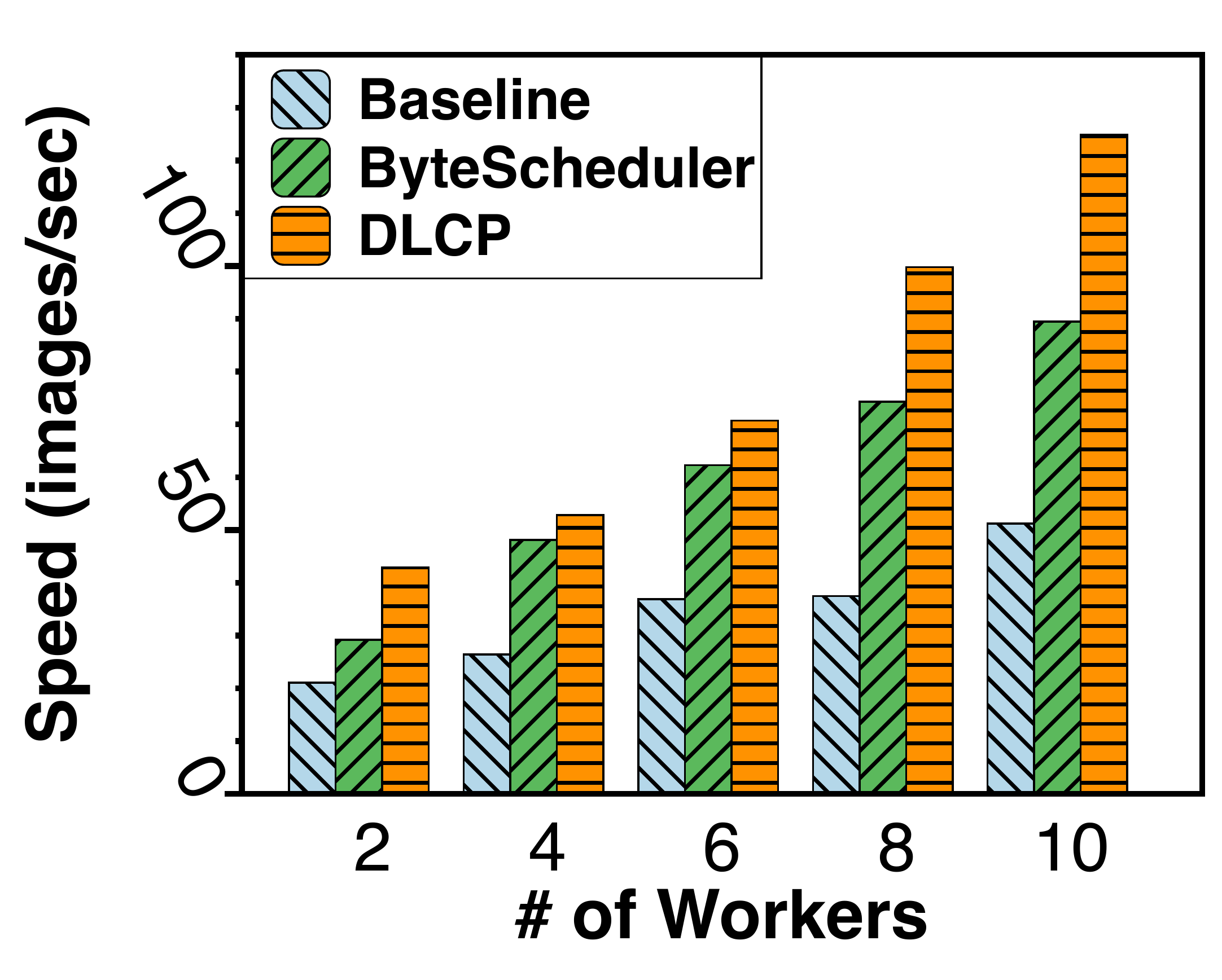}
    }
    \caption{Speedup with Different Frameworks.}
    \label{fig:testbed:frameworks}
\end{figure*}

\begin{figure}[htbp!]
    \centering
    \subfigure[Ring All-reduce, ResNet50]{
	\includegraphics[width=0.47\linewidth]{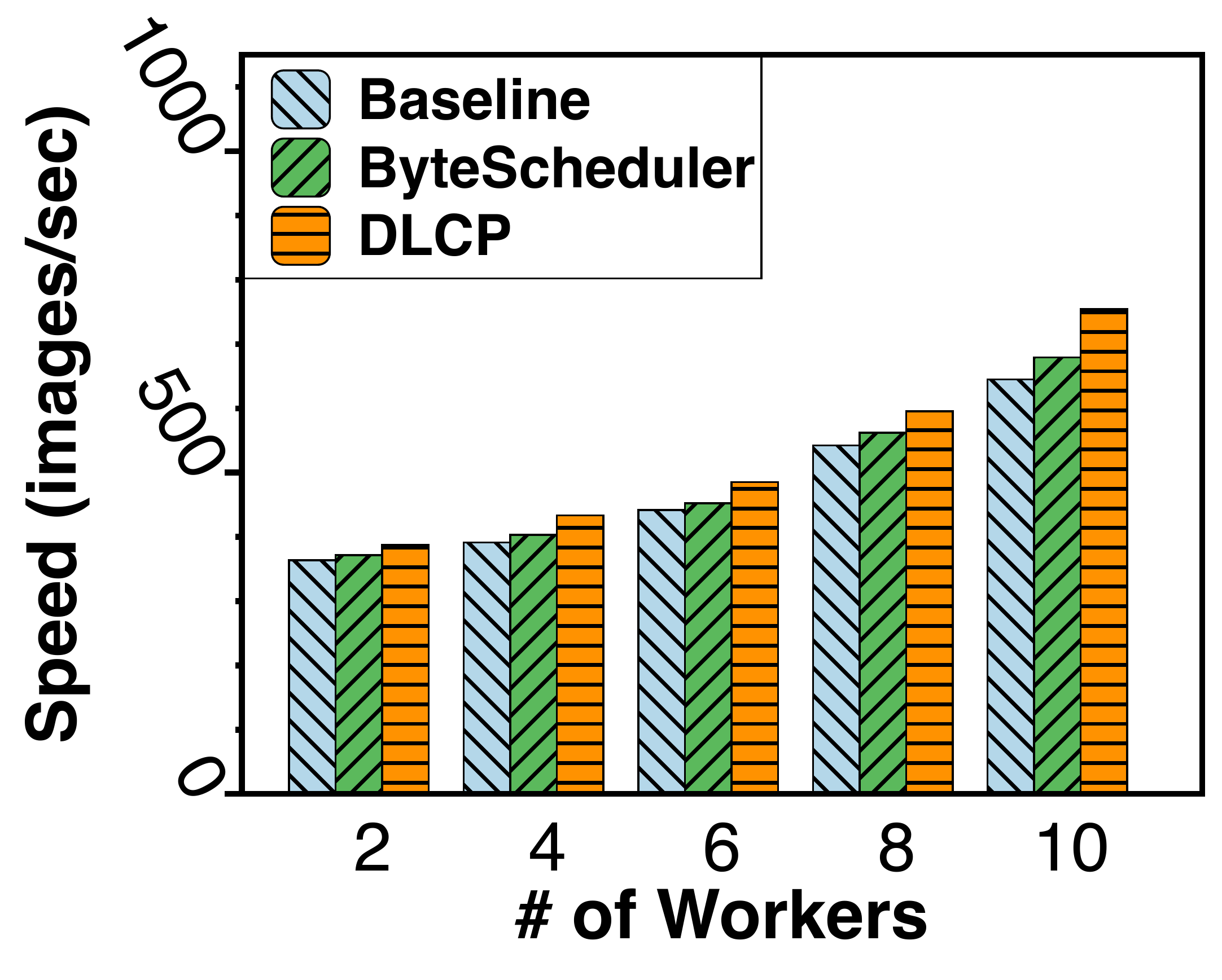}
    }
    \subfigure[Ring All-reduce, VGG16]{
	\includegraphics[width=0.47\linewidth]{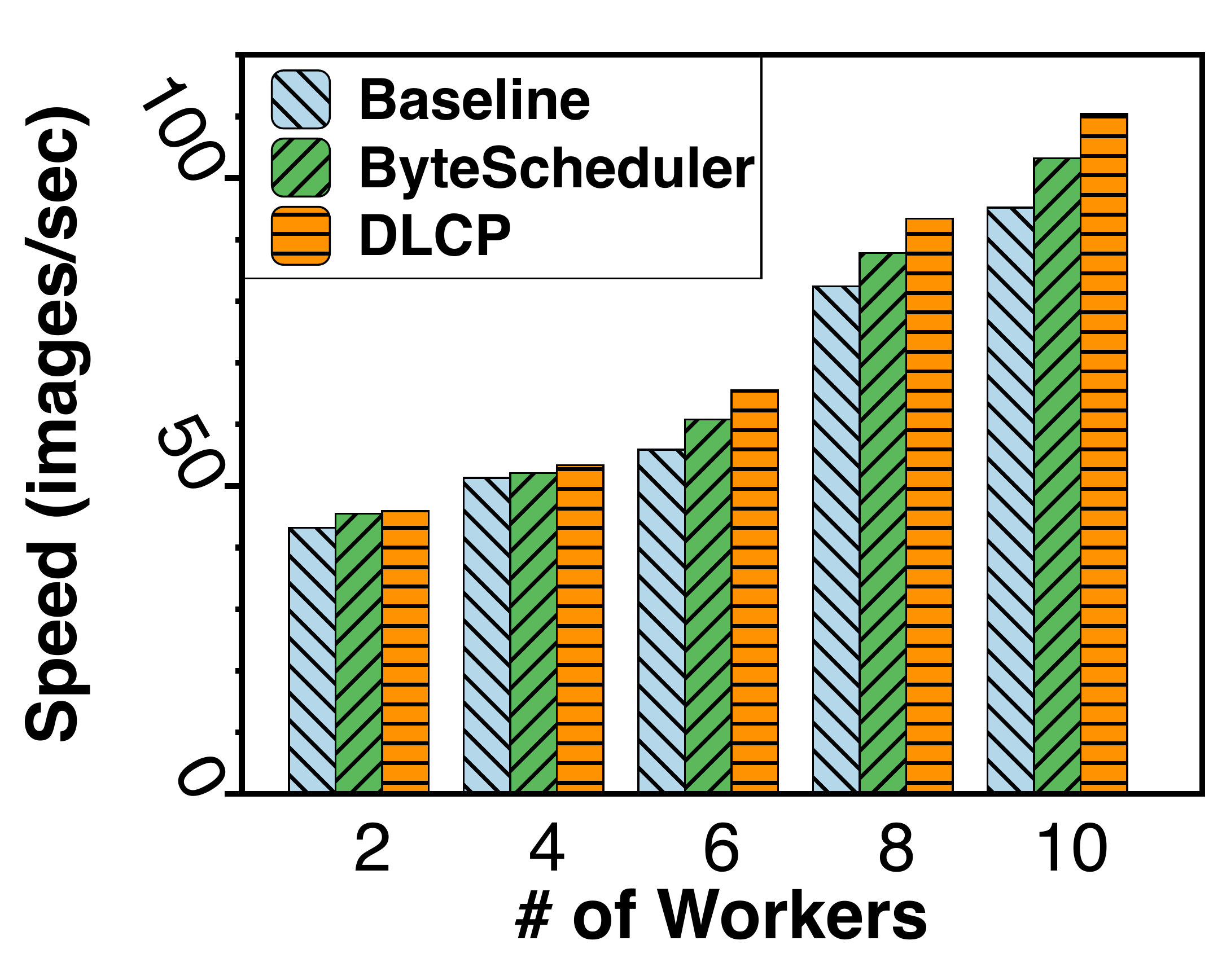}
    }
    \caption{Speedup with Ring All-reduce Synchronization Schemes.}
    \label{fig:testbed:psring}
\end{figure}

\subsection{Testbed Experiments}\label{subsec:experiment}

We integrate \sys into ByteScheduler, a state-of-the-art DNN scheduler supporting Tensorflow, PyTorch and MXNet, and evaluate its end-to-end performance in a small-scale testbed. 


\subsubsection{Experimental Setup}

\parab{Testbed:} Our testbed has 5 physical machines (each with 2 Tesla V100 GPUs, 20 CPU cores, 128GB memory), and 4 Mellanox SN2100 switches running Onyx 3.7.1134 operating system. We put two GPUs of one physical machine into two separated dockers with different network interfaces, therefore we get 10 logical nodes. We build a leaf spine topology with two core switches and two top-of-rack (ToR) switches. Each ToR switch is connected to five server nodes using 10Gbps and two core switches using 25Gbps links. 

\parab{Models and Dataset:} 
We use four models and two datasets in our experiments. Our models include three image classification tasks: VGG16\cite{VGG}, ResNet50\cite{resnet2016cvpr} and Inception-v3\cite{inception} training on the synthetic data with the same image size as imagenet\cite{imagenet}; and one for translation task: Transformer\cite{transformer} training on SQuAD\cite{Rajpurkar_2016}. We run experiments on three popular machine learning frameworks: TensorFlow, PyTorch, MXNet by using two different parameter synchronization schemes, PS and Ring All-reduce. 

\parab{Parameter settings:} The batch sizes of VGG16, ResNet50, Inception-v3 and Transformer are 16, 32, 32, and 10 samples per GPU. Switches have 4MB memory pool shared by all ports, and 8 queues for each port. Default transport protocol is TCP CUBIC\cite{tcpcubic}, RTOmin is 5ms and initial window size is 10. We set loss-tolerant bound to 10\% for \sys.

\parab{Baselines and metrics:} We compare \sys with the vanilla ML frameworks (baseline, the aforementioned three frameworks), and ByteScheduler with TCP as default transport protocol. For image classification models, we use the number of images processed in one second as the speed metric, and for transformer, we use the number of examples\cite{Rajpurkar_2016}.




\subsubsection{Overall results}
We test \sys across different DNN models, frameworks and synchronization schemes. \sys achieves speedup without affecting the convergence accuracy or increase the convergence round. Here, we show the speedup of each experiment.

\parab{Speedup under different DNN models:}
We compare \sys with four different models in PS architecture by using TensorFlow as the implementation framework. From the experiment results, we get the following three observations. 
First, \sys performs the best under all models. Figure~\ref{fig:testbed:models} shows that \sys outperforms ByteScheduler by 45.7\%-73.4\%, and baseline by 104.6\%-368.5\%, across the four models. The main reason is that default reliable transport is sensitive to packet losses, which may trigger timeouts and causes millisecond-level delay. During the training, we observed about 0.53\% packet losses from the buffer counting function provide by our switch\cite{mellanox}. 
Second, the improvement of \sys is more significant with the number of workers increases. As we can see, \sys outperforms ByteScheduler from 29.0\% to 36.2\% when the number of workers is 2, while it is 45.7\%-73.4\% with the number of workers is 10. This is expected because the network burden is higher and hence packet losses are more frequent. Third, \sys achieves more speedup in ResNet50 than other models. From figure~\ref{fig:testbed:models}(a), \sys achieves up to 73.4\% speedup in ResNet50, more than the speedup of any other models. This is because ResNet50 has much more layers and smaller layer size, thus generating much more small flows, which are susceptible to tail packet drops.


\parab{Speedup under different frameworks:}
We implement \sys with different machine learning frameworks. Figure~\ref{fig:testbed:frameworks} shows the training speed of ResNet50 and VGG16 in PS architecture with PyTorch and MXNet implementations. Notice that there are no default PS implementation in PyTorch, so we implement the PS based on the PyTorch distributed package\cite{PyTorch_dist}. From the figures, \sys outperforms ByteScheduler by up to 84.3\%/79.9\% in PyTorch and MXNet. That indicates \sys can achieve significant performance improvement with various frameworks.

\parab{Speedup under different synchronization schemes:}
We also evaluate the performance of \sys in Ring All-reduce architecture with ResNet50 and VGG16 (Figure~\ref{fig:testbed:psring}).
In both cases, we can find that \sys achieves higher speedup in PS architecture than Ring All-reduce. In PS, \sys achieves 84.3\%/45.7\% improvement for ResNet50 and VGG16, respectively. While the numbers are only 11\% and 7\% in Ring All-reduce\footnote{Note that ByteScheduler does not support ring communication directly, we put some effort into making it adopt to Ring All-reduce. Therefore, implementation overhead (e.g. data copy in framework) introduces additional overhead, and slowdown the speedup in ring.}. The reason is that: Packet loss is rare in Ring All-reduce, therefore Ring All-reduce is free from the long tail latency caused by timeout. Meanwhile, \sys still achieve improvement in Ring All-reduce, due to the fine-granularity layer level scheduling and reduction of data volume in transmission with bounded loss tolerance. In addition, we observe that \sys achieve more improvement in ResNet50 in PS while less in Ring All-reduce, the reason is that tensors in ResNet50 are relatively small, therefore more easy to trigger timeout in PS. For VGG16, flows are larger, which magnifies the shortcomings of single path.

\subsection{Large-scale Simulations}
Next, we use NS3 simulator to evaluate \sys's performance on large-scale networks. In simulations, we simulate empirical traffic based on communications patterns observed in real DNN training workloads. 




\subsubsection{Simulation Setup}
\parab{Topology:} We chose a leaf-spine topology, which has 4 core switches and 9 ToR switches, each rack has 16 hosts. Each ToR switch is connected to 16 hosts using 100Gbps links and 4 core switches using $4\times 100$Gbps links. Base round-trip time between two servers(4 hops) is $85.2\mu s$. Switch queue buffer size is 512KB per port.

\parab{Traffic:} We simulate traffic patterns of training ResNet50 and InceptionV3 under Parameter Server, workers are colocated with parameter servers and the ratio of their numbers is 4:1. We fetch computation completion time and size of each tensor from our testbed. Workers assign tensors randomly to different parameter servers, when tensor size is larger than a threshold(in our setting, 4MB), it is divided evenly to all parameter servers.

\parab{Baseline and metrics:} We use DCTCP\cite{dctcp} as the baseline, since it is widely used in many production data centers. We also compare with PIAS\cite{pias}, pFabric\cite{pfabric2013sigcomm}. TCP initial window is 10, and ECN marking threshold is 65 packets. RTOmin\footnote{RTOmin: TCP minimum retransmission timeout, we will analyze this parameter in \S\ref{sec:eval:deep}}\cite{vasudevan2009safe, bai2020one} is 10ms. By default, DupACKs is 3 for DCTCP and PIAS, and DelayAck is disabled. DCTCP and PIAS use per-flow ECMP while pFabric uses per-packet ECMP. We measure both average flow completion time(FCT) and tail FCT.

\begin{figure}[htbp!]
    \centering
    \subfigure[ResNet50, Average FCT]{
	\includegraphics[width=.47\linewidth]{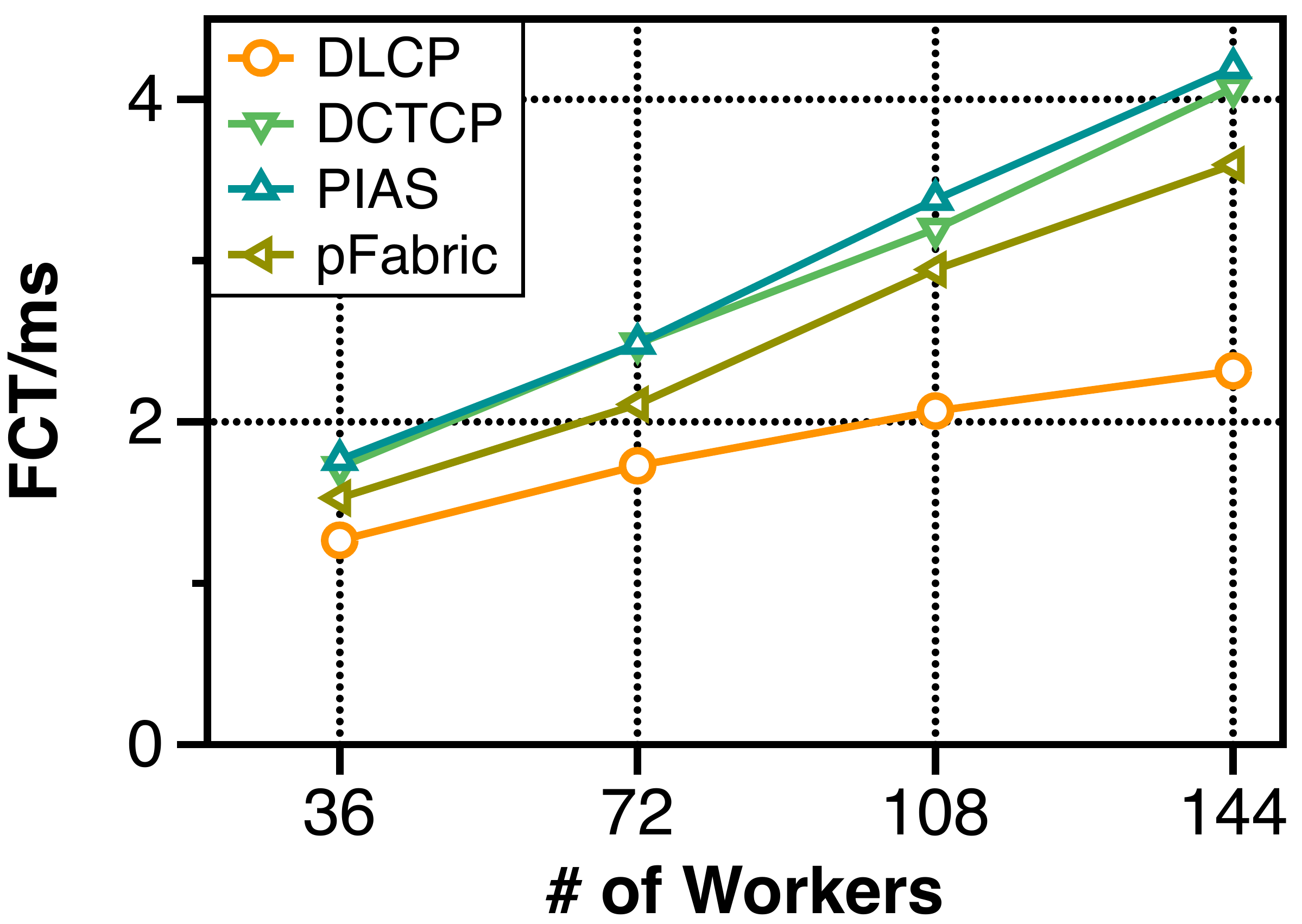}
    }   
    \subfigure[ResNet50, Tail FCT]{
    	\includegraphics[width=.47\linewidth]{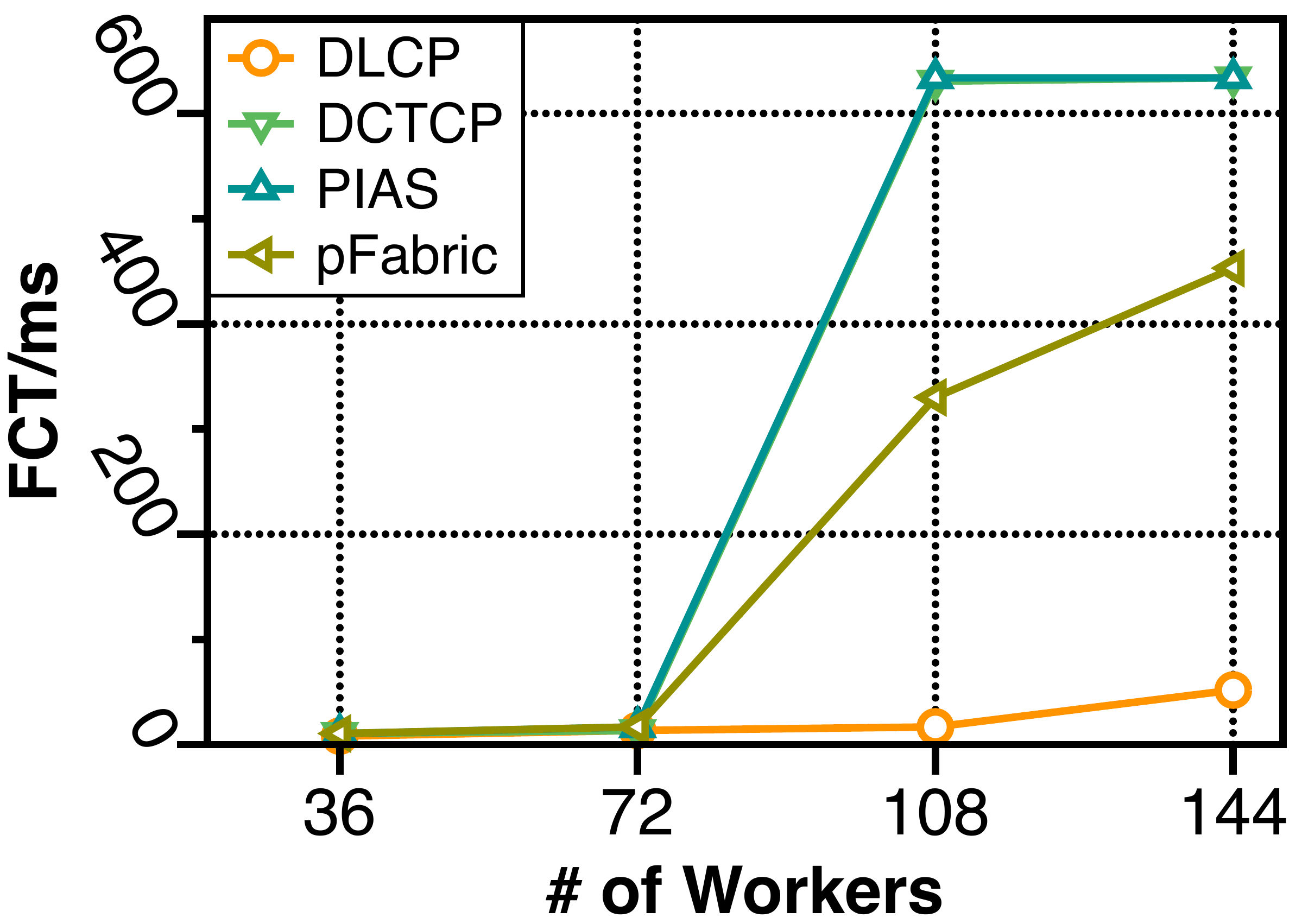}
    } 
    \quad
    \subfigure[InceptionV3, Average FCT]{
	\includegraphics[width=.47\linewidth]{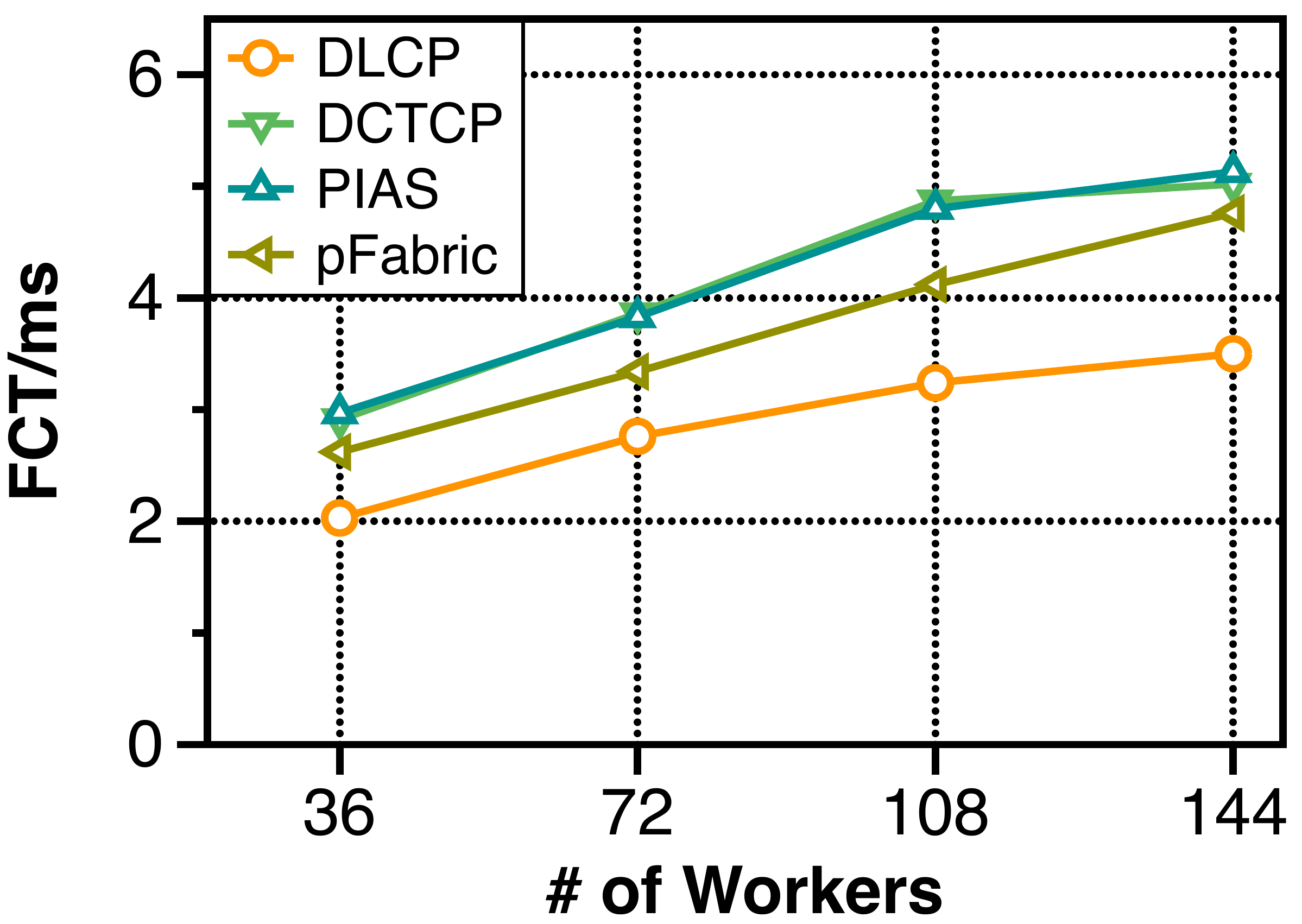}
    }  
    \subfigure[InceptionV3, Tail FCT]{
        \includegraphics[width=.47\linewidth]{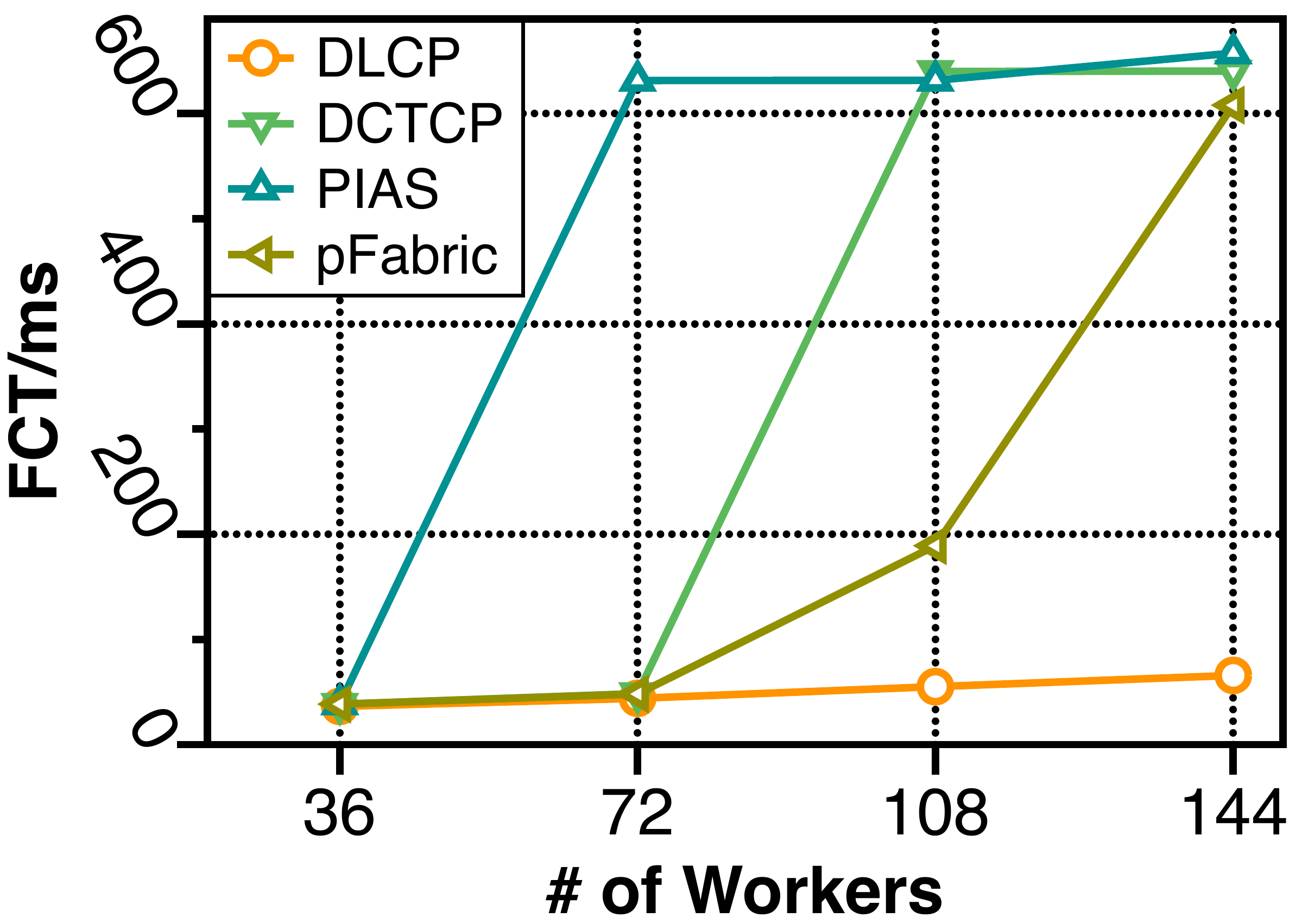}
    }
    \caption{Result of Large-scale Simulations.}
    \label{fig:sim:servers}
\end{figure}

\subsubsection{Results}

Figure~\ref{fig:sim:servers} gives both average/tail FCT for ResNet50 and InceptionV3 with various scales. Y-axis indicates the flow completion time, and X-axis indicates the total number of workers (workers and servers are evenly located on each rack). 
In general, \sys delivers the best performance. For ResNet50, \sys is up to 43.1\%/44.8\%/35.5\% lower average FCT and 91.8\%/91.8\%/88.6\% lower tail FCT compared to DCTCP, PIAS, and pFabric. For InceptionV3, \sys reduces average FCT by up to 30.3\%/31.8\%/26.5\% and tail FCT up to 89.7\%/90.0\%/89.2\%, respectively.
According to the above results, we make the following three observations: 
\begin{icompact}

\item \textit{\sys preforms the best in various settings.} 
\sys achieves the best performance in all workloads and network scales, especially in large scale networks. The reason is that other algorithms suffer from packet retransmission to keep reliability. In PS, multi-workers currently send gradients to the same server, which causes incast\cite{incast} happens and leads to packet loss. 

\item \textit{\sys significantly reduces tail FCT.} 
Compared to average FCT, \sys reduces the tail more significantly. The reason is that, retransmission timeout greatly increases the tail of other algorithms, while \sys tolerates packet loss and free of retransmission timeout.  



\item \textit{The speedup of \sys is more notable as the scale increases.} In general, \sys reduces FCT more significant in the large scale. The reason is that, as the scale increases, the prior solutions suffer from more packet loss, therefore cause performance degradation, while \sys is immune to packet loss with the loss tolerance design.

\end{icompact}

\subsection{Deep Dive}\label{sec:eval:deep}

 \begin{figure}[htbp]
     \centering
     \subfigure[Accuracy vs. epoch]{
         \includegraphics[width=0.47\linewidth]{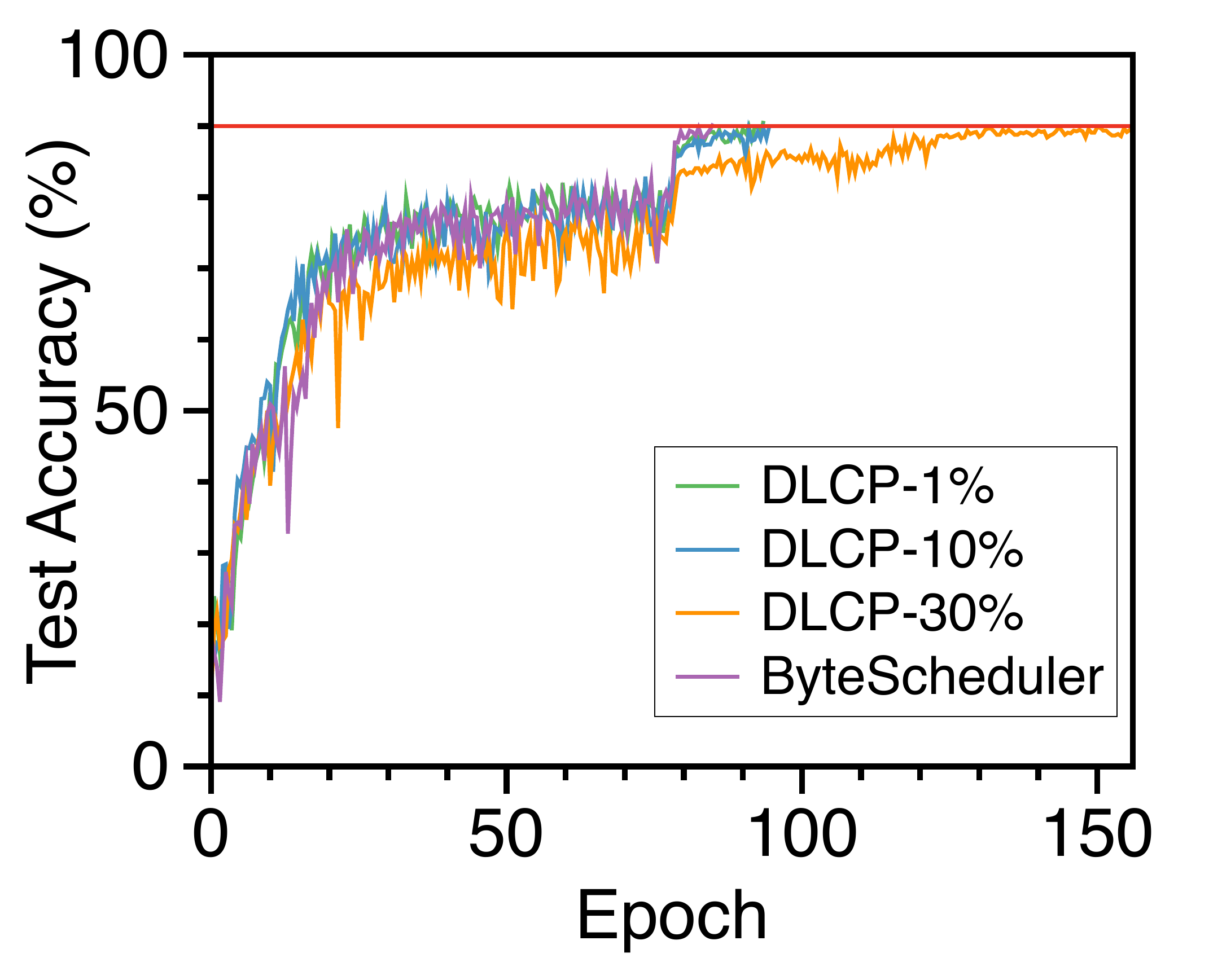}
     }
     \subfigure[Accuracy vs. time]{
 	\includegraphics[width=0.47\linewidth]{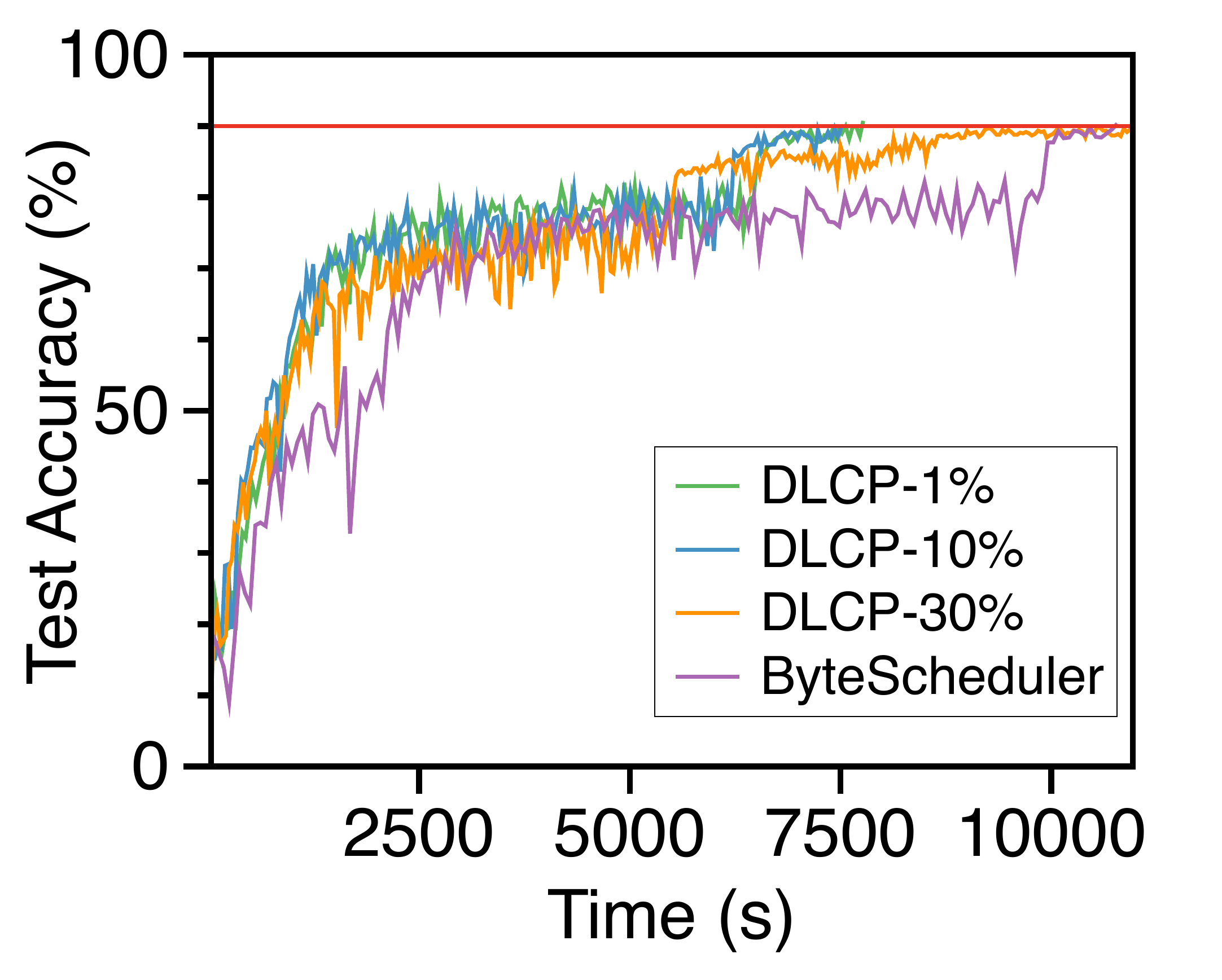}
 	 }
     \caption{Impact of Loss-tolerant Bound.}
     \label{fig:testbed:dd:loss}
 \end{figure}

\parab{Impact of loss-tolerant bound:}
We measure convergence and speedup of \sys with different loss-tolerant bounds (1\%, 10\%, 30\%) of ResNet50 on Cifar10\cite{cifar}.
Figure~\ref{fig:testbed:dd:loss}(a) shows curves of test accuracy vs. epoch, for DLCP with 1\% and 10\% loss-tolerant bound, the curves are almost in line with the benchmark (ByteScheduler~\cite{bytescheduler}), and for DLCP with 30\% loss-tolerant bound, with more training epochs, it can eventually reach the same test accuracy. This highlights the fact that loss-tolerant bound of \sys can be set to 10\% without affecting model convergence. Figure~\ref{fig:testbed:dd:loss}(b) shows test accuracy over time. As we can see, compare to ByteScheduler, \sys converges faster under all loss-tolerant bounds. Meanwhile, we find that 1\% and 10\% loss-tolerant bound take almost the same time to converge. This indicates that we do not need to fine-tune the loss-tolerant bound to achieve the state of art performance.

\begin{figure}[htbp]
    \centering
    \subfigure[ResNet50, RTOmin=1ms]{
	\includegraphics[width=.47\linewidth]{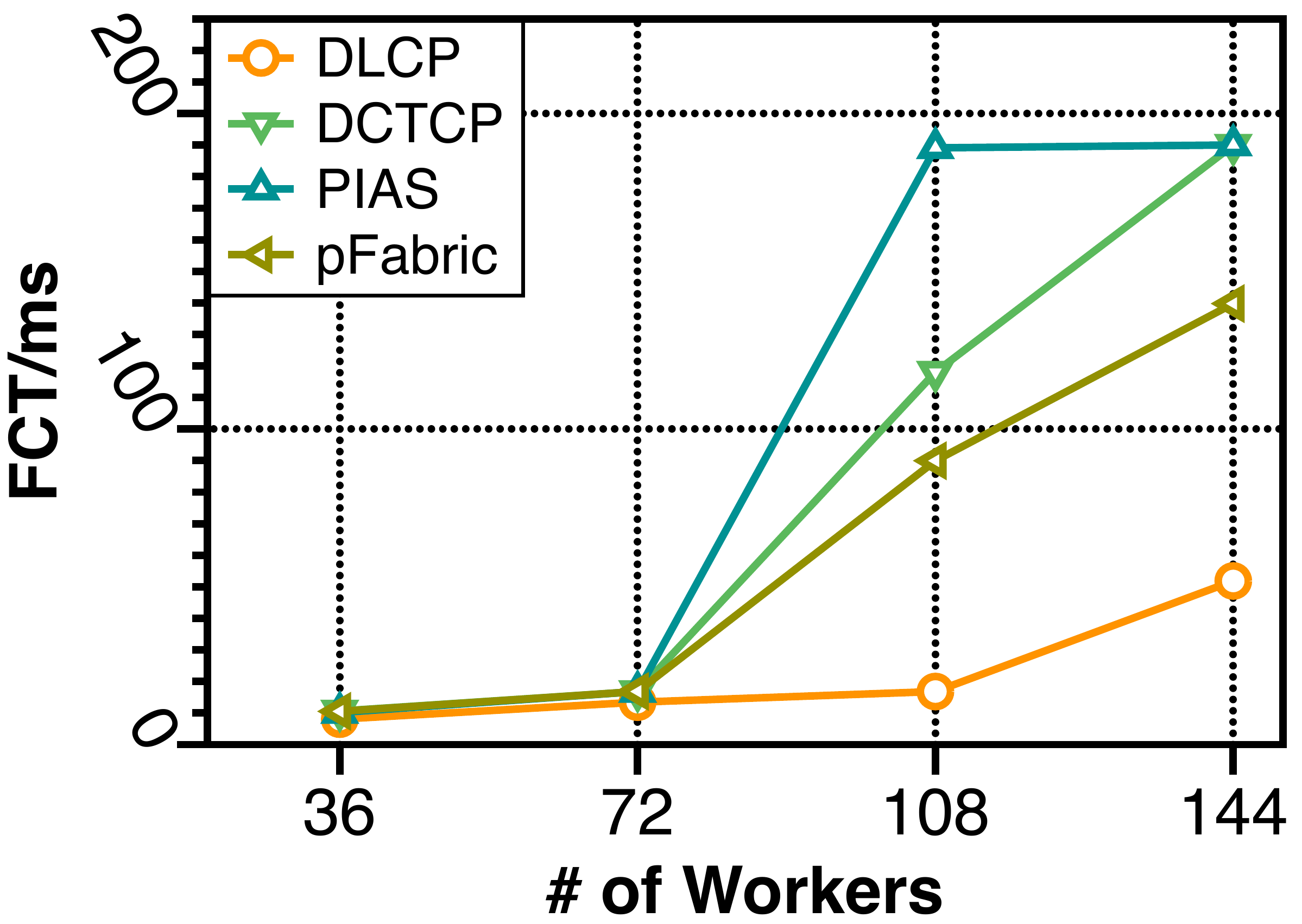}
    }
    \subfigure[InceptionV3, RTOmin=1ms]{
        \includegraphics[width=.47\linewidth]{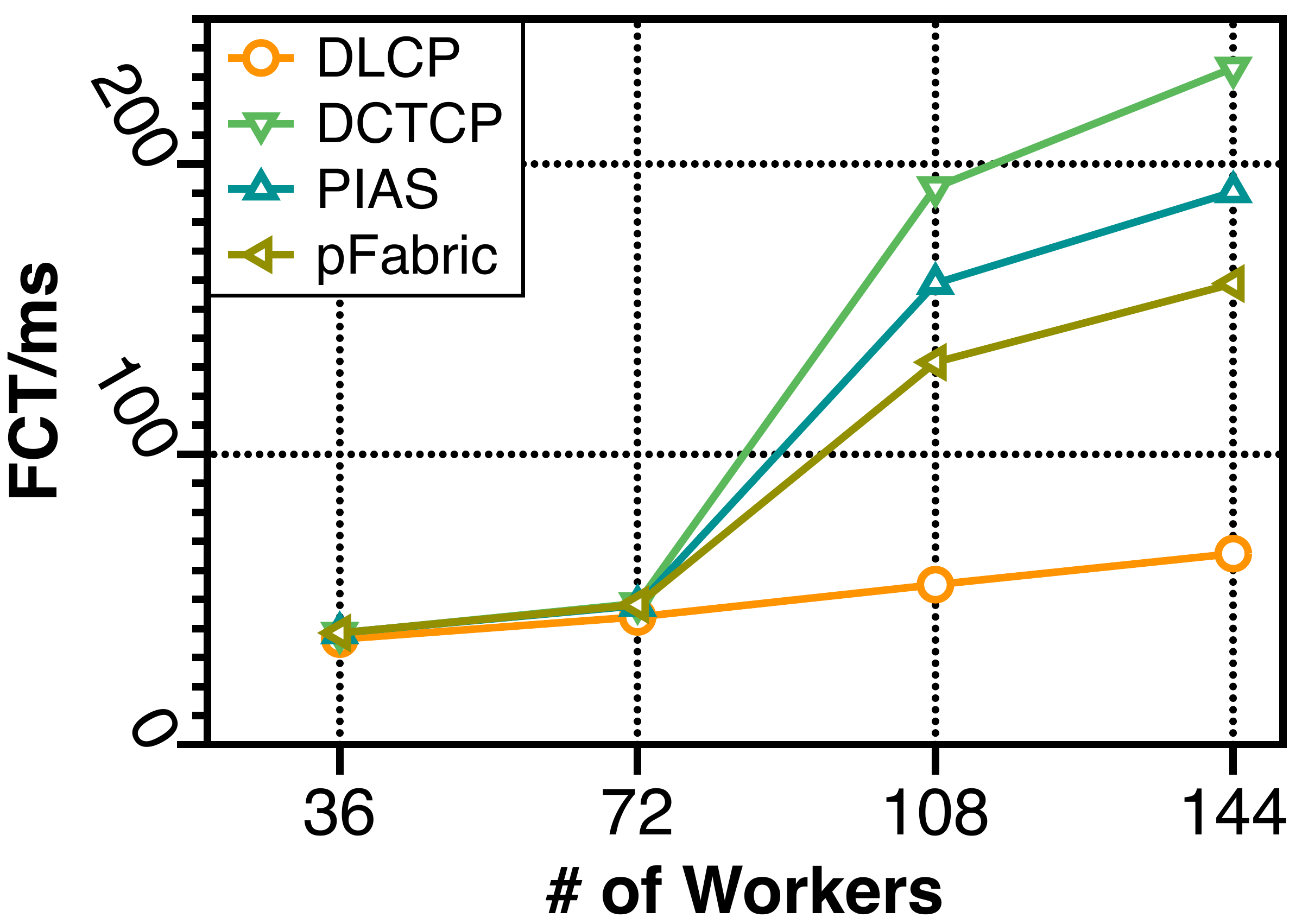}
    }
    \quad
    \subfigure[ResNet50, RTOmin=5ms]{
	\includegraphics[width=.47\linewidth]{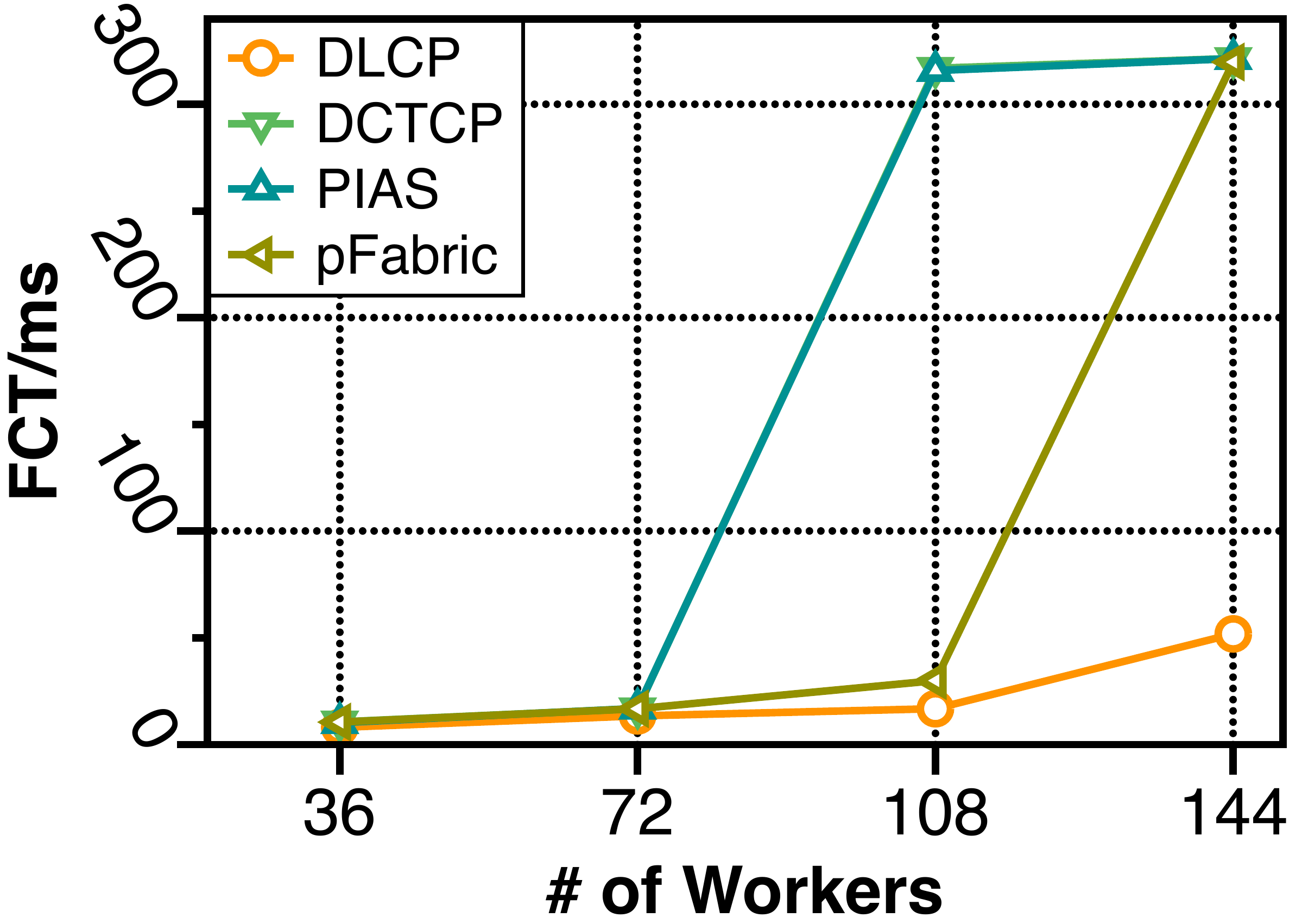}
    } 
    \subfigure[InceptionV3, RTOmin=5ms]{
    	\includegraphics[width=.47\linewidth]{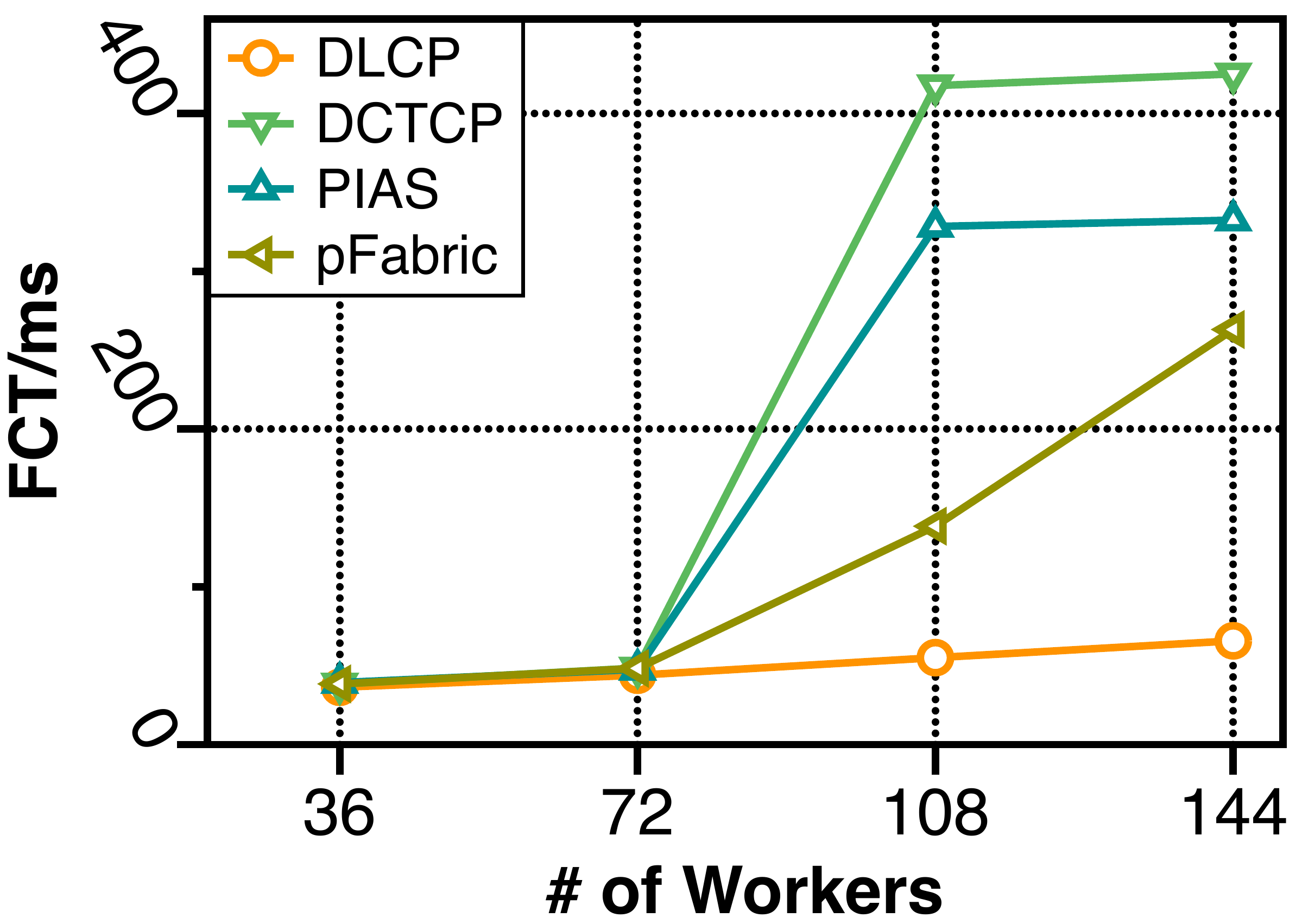}
    }
    \caption{Tail FCT under Different RTOmins.}
    \label{fig:sim:rtomin}
\end{figure}

\parab{Impact of retransmission timeout:}
Although retransmission timeout does not impact \sys, it may influence other compared algorithms. To exclude the impact of RTO setting, we perform a simulation experiment with two smaller RTOmin values. Figure~\ref{fig:sim:rtomin} shows the tail FCT under ResNet50 and InceptionV3 with two different RTOmin settings, \sys still reduces tail FCT by up to 83.9\%/72.7\% with RTOmin equal to 5ms, and 84.5\%/71.8\% with RTOmin equal to 1ms. The reason is that although simply reduced RTOmin saves time from waiting for retransmission, it introduces other problems like spurious retransmission\cite{ludwig2000eifel}. Meanwhile, previous work\cite{vasudevan2009safe} shows that many factors prevent us from reducing RTOmin, such as low-resolution timers and delayed acknowledgements\cite{delayack}. \sys, instead, can tolerate packet loss without retransmission, therefore is free of the above drawbacks of changing retransmission timeout setting.

\section{Related Work}\label{sec:related}


\parab{Optimizing communication in DNN training.}
Besides the closely related works discussed above, some other methods have been proposed to improve the communication of DNN training. For example, RDMA~\cite{rdma, yi2017towards, hu2016deadlocks} and NCCL~\cite{nccl} provide higher bandwidth between workers to speedup tensor transmission. Works like BlueConnect~\cite{blueconnect} and PLink~\cite{plink} design novel communication patterns with network topology awareness for gradient synchronization process at each iteration for better performance and robustness. GPipe~\cite{gpipe} and PipeDream~\cite{pipedream} overlap communication with computation in model parallelism context. Traditional works like flow scheduling ~\cite{pias, bai2014pias, bai2017pias, chen2018auto} and coflow scheduling~\cite{susanto2016stream, chowdhury2015efficient, zhang2016coda} can also optimize communication by minimizing flow (coflow) completion time.  Note that these works are orthogonal to \sys.

Some works like ASP~\cite{hogwild2011nips} and SSP~\cite{petuum} propose synchronization algorithms to relax synchronization requirements. A recent work, SwitchML~\cite{switchml} leverages in-network aggregation to reduce the communication overhead in network. Although \sys cannot directly integrate with these algorithms, its three core ideas can still be used to do further optimization, we leave it to future work.


\parab{Load balancing.}
ECMP~\cite{ecmp}, MPTCP~\cite{mptcp}, Conga~\cite{conga}, Hermes~\cite{hermes} and Letflow~\cite{letflow} conducts load balancing at flow or sub-flow level, whereas \sys implements a more fine-grained per-packet load balancing solution. And compared to other per-packet load balancing schemes~\cite{drb,drill,ndp}, \sys leverages domain specific knowledge of distributed DNN training, which greatly simplifies the overall design by tolerating packet drops and reordering.

\section{Conclusion}\label{sec:conclusion}
This paper presented \sys, a novel solution exploiting the domain-specific properties of deep learning to optimize communication overhead of DNN training in a fine-grained manner. At its heart, \sys comprises of three key innovations beyond prior work: 1) cutting tail communication latency via bounded-loss tolerant data transmission, 2) maintaining training efficiency via DNN-aware priority queueing and dropping, and 3) performing per-packet load balancing based on inter-packet order-independency. We have implemented \sys with commodity switches, integrated it with various training frameworks including TensorFlow, MXNet and PyTorch, and deployed it in our small-scale testbed with 10 Nvidia V100 GPUs. Our testbed experiments and large-scale simulations demonstrated great potential of \sys: it delivers up to $84.3\%$ additional training acceleration over prior solutions.

\bibliographystyle{plain}
\bibliography{mltref.bib}
\clearpage
\appendixpage
\appendix
\section{Convergence Proof of \sys}\label{sec:analysis:converge}

In this section, we give a convergence proof for distributed machine learning with priority dropping mechanism based on gradient magnitude level.  We use the notations as the table~\ref{sec:analysis:notation} shows.


\begin{table}[htbp]
\caption{Definitions and notations}
\begin{tabularx}{0.48\textwidth}{p{0.07\textwidth}X}
\toprule
  $\lVert\cdot\rVert$ & $l_2$ norm for vectors \\
  $\lVert\cdot\rVert_F$ & the Frobenius form of matrices \\
  $n$ & number of workers \\
  $m$ & number of servers \\
  $\gamma$ & model learning rate \\
  $p$ & packet dropping ratio \\
\bottomrule
\end{tabularx}
\label{sec:analysis:notation}
\end{table}

\parab{The distributed machine learning model}
\cite{ankit} has proved the comparable convergence rate of distributed learning over unreliable network with independent and equivalent packet drop probability $p$ for each message. Based on their Reliable Parameter Server (RPS) algorithm, we consider the parameter server model over our \sys with priority dropping mechanism: the packet drop probabilities differ in gradients of different magnitudes. 

The distributed optimization problem is defined as:
\begin{equation}
\min_{\vec{x}}{f(\vec{x})=\frac{1}{n}\sum_{i=1}^{n}f_i(\vec{x})},
\end{equation}

where $n$ is the number of workers, $f_i(\vec{x})=\mathbb{E}_{\xi~D_i}F_i(\vec{x},\xi)$ represents the expected loss function $F$ over $D_i$, the local data distribution of worker $i$.
At each iteration, every worker performs SGD on a random chosen subset of dataset $D_t^{(i)}$. $$G_t^{(i)} = \nabla F_i\left(X_t^{(i)}, D_t^{(i)}\right).$$
$X_t^{(i)}$, $G_t^{(i)}$ and $D_t^{(i)}$ denotes the model weights, generated gradients and training data of worker $i$ at iteration $t$ respectively. 
 
Before sending the gradients, every worker $i$ divides the gradients into $m$ equal blocks:
$$G_t^{(i)} = \left( (G_t^{(i, 1)})^\intercal, (G_t^{(i, 2)})^\intercal, \dots, (G_t^{(i, m)})^\intercal \right) .$$
When sending  gradients $G_t^{(i)}$, some blocks may be dropped because of the networking condition and priority dropping. For each blocks, the gradients on every workers are collected and averaged by parameter server:
$$ \widetilde{G}^{j}_t=\frac{1}{|N_t^{(j)}|}\sum_{i\in N_t^{(j)}}G^{(i,j)}_t,  $$
where $\widetilde{G}^{j}_t$ denotes the averaged gradients of block $j$ at iteration $t$, and $N_t^{(j)}$ denotes the number of workers whose blocks $j$ are successfully averaged at iteration $t$.

After averaging gradients, the parameter server updates the corresponding weight block using SGD algorithm and returns them back to each workers for their local updates. For workers that fail to receive the averaged block, they just use the original gradients. Formally, the updated gradients on worker $i$ is
$$ X^{(i)}_{t+1}=\left( (X_{t+1}^{(i, 1)})^\intercal, (X_{t+1}^{(i, 2)})^\intercal, \dots, (X_{t+1}^{(i, m)})^\intercal \right),$$
where 
$$ X_{t+1}^{(i, j)}= \left\{     \begin{aligned}  &  X_{t}^{(i, j)}- \gamma  \widetilde{G}^{j}_t   , & i \in \widetilde{N}_t^{(j)} \\   &  X_{t}^{(i, j)}- \gamma  G^{(i ,j)}_t  , & i \notin \widetilde{N}_t^{(j)}.     \end{aligned}     \right.$$
$\widetilde{N}_t^{(j)}$ denotes the set of workers to which the averaged block $j$ is successfully sent at iteration $t$.

For the algorithm, we make the following assumptions commonly used for analyzing stochastic optimization algorithms \cite{ankit,lian2015asynchronous}.
\begin{assumption}We make the following commonly used assumptions:
	\begin{enumerate}
		\item \textbf{Lipschitzian gradient}: The gradient function $\nabla f_i(\cdot) $ is L-Lipschitian, which means 
		$$  \| \nabla f_i(\vec{x}) - \nabla f_i(\vec{y})\|\leq L\| \vec{x}-\vec{y}\|  $$
		\item \textbf{Bounded gradient}: The variance of stochastic gradient is bounded for every worker $i$ and any $\vec{x}$.
		$$ \begin{aligned}\mathbb{E}_{\xi ~ D_i}\| \nabla F_i(\vec{x};\xi) -\nabla f_i(\vec{x})\|^2 \leq \sigma^2, & \forall i, \forall \vec{x} \\ \frac{1}{n}\sum_{i=1}^{n}\|\nabla f_i(\vec{x})-\nabla f(\vec{x}) \|^2 \leq \xi^2, & \forall i, \forall \vec{x},\end{aligned}$$
		\item \textbf{Start from 0}: For simplicity, we assume $X_1=0$ w.l.o.g.
	\end{enumerate}
\end{assumption}

With arbitrary packet dropping policy, the updated gradients on each worker can always be represented as the linear combination of local gradients. 
$$  X^{(i,j)}_{t+1}-X^{(i,j)}_{t}=G^{(\cdot ,j)}_t W^{(j)}_t, $$
where 
$$ G^{(\cdot ,j)}_t := \left( (G_{t}^{(1, j)})^\intercal, (G_{t}^{(2, j)})^\intercal, \dots, (G_{t}^{(i, j)})^\intercal \right). $$
$W^{(j)}_t$ is the coefficient matrix. And $\left[W^{(j)}_t\right]_{m,k}$ denotes the coefficient of worker $m$'s gradients received by worker $k$ after one update step. $\left[W^{(j)}_t\right]_{m,k}=0$ means worker $m$'s gradient block $j$ is not received by worker $k$, which may be dropped either before or after the averaging during the communication with the parameter server. 

\cite{ankit} shows $W^{(j)}_t$ satisfies the following properties under uniformly random dropping environment:
\begin{gather}\label{eq:property}
\mathbb{E}[W]=\alpha_1 I_n+(1 - \alpha_1)A_n  \\
\mathbb{E}[W^{(j)}_t W^{(j)^\intercal}_t] \alpha_1 I_n+(1 - \alpha_1)A_n \\
\mathbb{E}[W^{(j)}_t A_n W^{(j)^\intercal}_t] =\alpha_2 I_n + (1-\alpha_2)A_n
\end{gather}
for some constants $\alpha_1$ and $\alpha_2$ satisfying $0<\alpha_2<\alpha_1<1$. While \cite{ankit} considers the algorithm where workers perform the averaging operation, the properties also hold for dedicated parameter server setting. Also,
as \sys adopts priority dropping mechanism, $(\alpha^{(j,t)}_1,\alpha^{(j,t)}_2)$ varies in different blocks $j$ and iterations $t$. To adopt the convergence proof in \cite{ankit} for \sys, we use $\alpha_{1_{max}}, \alpha_{2_{max}}$ instead, which denotes the maximum value of $\max_{j,t}{\alpha^{(j,t)}_1}$ and $\max_{j,t}{\alpha^{(j,t)}_2}$ across all workers and iterations and preserve the validity of the proof. Thus we can get the following theorem:
\begin{theorem}(Convergence of \sys).
Under Assumption 1, choosing learning rate $\gamma$ to be small enough satisfying $1-\frac{6L^2\gamma^2}{(1-\sqrt{\beta_{max}})^2}>0$, \sys have the following convergence rate:
\begin{equation}\begin{aligned}
\frac{1}{T} & \sum_{t=1}^{T}\left(\mathbb{E}\|\nabla f(\overline{\vec{x_t}}) \|^2 + (1-L\gamma)\mathbb{E}\|\overline{\nabla}f(X_t) \|^2\right) \\
& \leq  \frac{2f(\vec{0})-2f(\vec{x^*})}{\gamma T} + \frac{\gamma L \sigma^2}{n} + 4\alpha_{2_{max}}L\gamma (\sigma^2 + 3\xi^2) \\
& + \frac{2\alpha_{2_{max}}L\gamma + L^2\gamma^2 + 12\alpha_{2_{max}}L^3\gamma^3)\sigma^2C_1}{(1-\sqrt{\beta_{max}})^2} \\
& +  \frac{3(2\alpha_{2_{max}}L\gamma + L^2\gamma^2+12\alpha_{2_{max}}L^3\gamma^3)\xi^2C_1}{(1-\sqrt{\beta_{max}})^2,}
\end{aligned}
\end{equation}
where 
\begin{equation*}
\begin{aligned}
& \nabla f(\overline{\vec{x_t}}) = \nabla f(\frac{1}{n} \sum_{i=1}^{n}\vec{x}^{(i)}_t) \\
& \overline{\nabla}f(X_t) = \sum_{i=1}^{n}\nabla f_i(\vec{x}^{(i)}_t) \\
& \beta_{max} = \max_{j,t}(\alpha^{(j,t)}_1 - \alpha^{(j,t)}_2) \\
& C_1=\left(1-\frac{6L^2\gamma^2}{(1-\sqrt{\beta_{max}})^2} \right)^{-1}.
\end{aligned}
\end{equation*}
\end{theorem}
\newpage

It can be inferred from the definitions that $\beta=1$ if and only if the dropping probability of the gradient block is 1, which may cause the bound to be infinity. In \sys we can make the assumption that no gradient block has dropping probability equal to 1, since the magnitude of gradients varies among different iterations.

By choosing approriate learning rate $\gamma=\frac{(1-\sqrt{\beta_{max}})^2}{6L+3(\sigma+\xi)\sqrt{\alpha_{2_{max}}T}+\frac{\sigma\sqrt{T}}{\sqrt{n}}}$, we can get
\begin{equation}\label{eq:convergence}
\begin{aligned}
\frac{1}{T} & \sum_{t=1}^{T}\mathbb{E}\|\nabla f(\overline{\vec{x_t}}) \|^2 \leq \frac{(2f(\vec{0})-2f(\vec{x}^*)+L)\sigma}{\sqrt{nT}(1-\sqrt{\beta_{max}})} \\
& + \frac{(2f(\vec{0})-2f(\vec{x}^*)+L)(\sigma + \xi)}{1-\sqrt{\beta_{max}}}\sqrt{\frac{\alpha_{2_{max}}}{T}} \\
& + \frac{L^2(\sigma^2+\xi^2)}{(\frac{T}{n}+\alpha_{2_{max}}T)\sigma^2+\alpha_{2_{max}}T\xi^2}\\
& + \frac{(2f(\vec{0})-2f(\vec{x}^*)L}{T}
\end{aligned}
\end{equation}

We can see from the Equation \ref{eq:convergence} that the dominant term in the convergence rate ($O(1/\sqrt{nT})$) is consistent with previous works for both centralized SGD and decentralized SGD\cite{ankit,lian2017can}.

\enlargethispage{-6.5cm}

\section{Thresholds Setting of Priority Dropping}\label{sec:analysis:thres}

In this section, we analyze and give the approach to finding the optimal thresholds of priority dropping for minimizing the impact of packet loss on model convergence. We leverage the queueing theory\cite{queueingtheory} to derive optimal thresholds for any DNN model. From the analysis, we find that the optimal thresholds are determined by the size of each layer of the model and the impact of different gradients on convergence.

\begin{figure}
    \centering
    \includegraphics[width=1.0\linewidth]{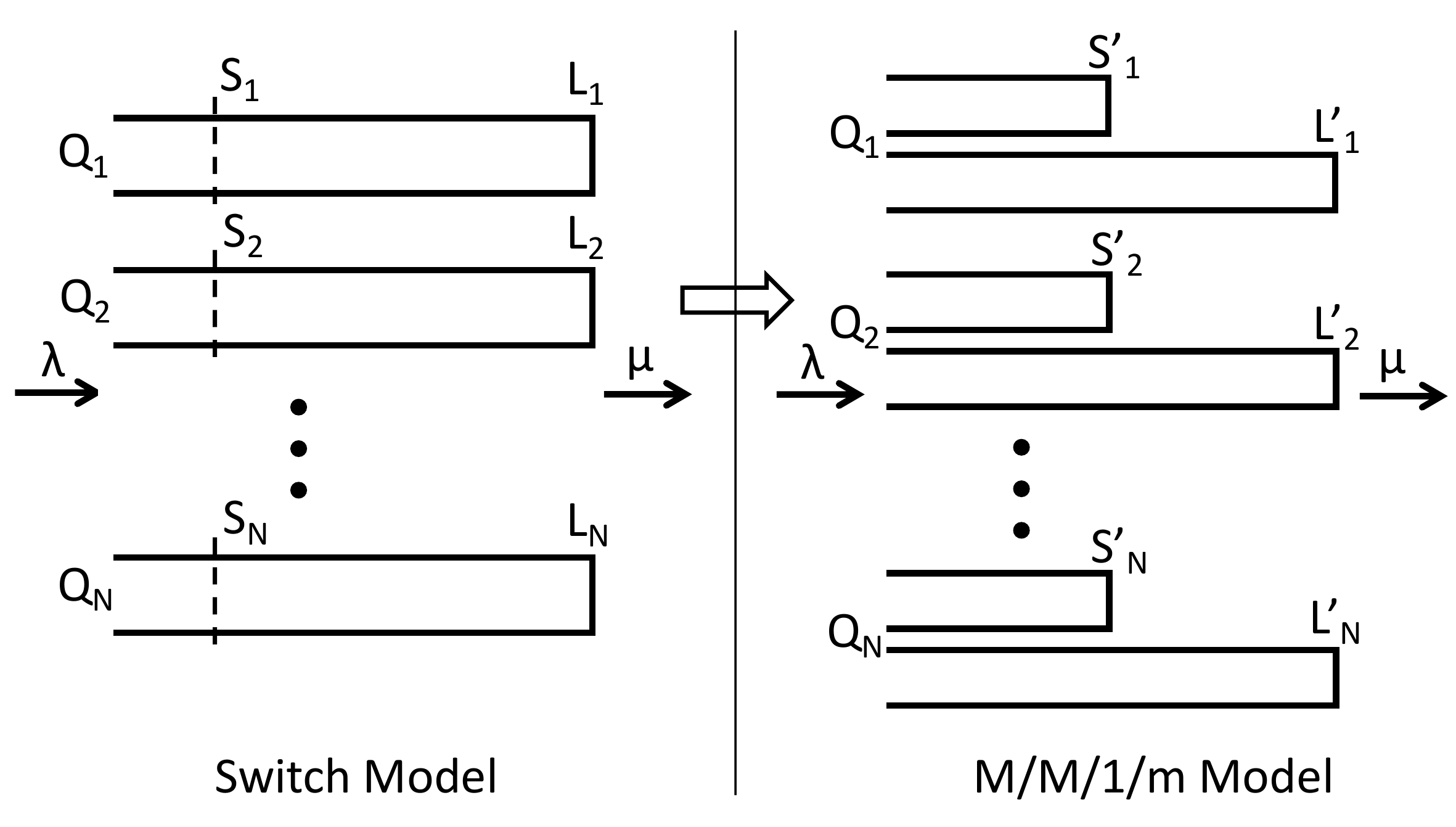}
    \caption{Problem formulation}
    \label{fig:analysis:thres}
\end{figure}

\parab{Problem formulation:} We assume the queue number is $N$ and the total buffer size is $B$, for the queue $i$, the ECN/RED threshold is $S_i$, and the queue size is $L_i$. Suppose the number of model layer is $M$, small gradient accounts for $\theta$, and $x\%$ loss of small/large gradients in queue i (layer $\vert \frac{M\cdot i}{N} \vert$ to $\vert \frac{M\cdot (i+1)}{N}-1 \vert$) costs $f^S(i)$/$f^L(i)$ additional convergence rounds, and the size of different layer in one model is $S^m(i), i\in [1,M]$, the total size of the model is $S = \sum_{k=1}^M S^M(k)$. Meanwhile, suppose the total packet arrival rate is $\lambda$ and the total service rate is $\mu$.

Figure~\ref{fig:analysis:thres} shows priority dropping and scheduling in \sys switch. All flows come at the arrival rate of $\lambda$, and enter the corresponding queues. If the queue length is larger than the threshold of small/large gradients, all small/large come to this queue will be discarded. Notice that the small and large gradients in one queue have different thresholds, to simplify the analysis and take the advantage of M/M/1/m Model in queueing theory, we split each queue into two, the first is for small gradients only and the second is for large gradients only. The dropping thresholds for each one are $S'_i$ and $L'_i$. We can easily represent $S_i$ and $L_i$ with $S'_i$ and $L'_i$: $S_i = S'_i / \theta, L_i = S'_i + L'_i$

Then, we deduce the value of $S'_i$ and $L'_i$. The arrival rate for one queue depends on the corresponding layers' packet arrival rate, for simplicity, we assume the rate is is proportional to the size of layer. Therefore, the arrival rate for Queue $i$ is $\frac{\lambda S^m(i)}{S}$, and for the small gradients' queue the value is $\theta \frac{\lambda S^m(i)}{S}$, for the large one, is $\left(1-\theta\right) \frac{\lambda S^m(i)}{S}$. The service rate for one queue is determined on its priority, it is serviced only when the higher priority queues are idle, for the highest priority queue $Q_1$, the service rate is $\mu_1 = \mu$, the idle time is $1 - \rho_1$, where $\rho_1 = \lambda /\mu$, for the queue $Q_2$, the service rate is $\mu_2 = (1-\rho_1)\mu$. Generally, the idle time for queue $Q_i$ is $1 - \rho_i$, where $\rho_i = \lambda_i /\mu_i$ and the service rate is $\mu_i = \Pi_{k=0}^{i-1}(1-\rho_k)\mu$, we can easily get the value of $\mu_i$ and $\rho_i$ from the above. Supposed the service rate for small/large gradients is proportional to the size, thus, the service rate for small/large gradients queue in queue $Q_i$ are $\mu^S_i = \theta \mu_i$ and $\mu^L_i = (1-\theta) \mu_i$ respectively. Therefore the idle time are $\rho_i^S = \lambda^S_i / \mu^S_i = \lambda_i / \mu_i = \rho_i = \rho^L_i$. 
Suppose the loss ratio for small/large gradients in queue $Q_i$ are $r^S_i$ and $r^L_i$, our goal is to minimize the impact of gradients' loss to model convergence, that is to find the optimal $S_i$ and $L_i$ to minimize the loss function $\sum_{i=1}^N \left(f^S(r^S_i)+f^L(r^L_i)\right)$, there are a lot of existing solutions to solve the optimization problem, e.g. gradient descent. Here, we only need to calculate the value of $r^S_i$ and $r^L_i$. In fact, for each queue, we can treat it as a typical M/M/1/m model in queueing theory, specially, one FIFO queue with finite capacity. Previous work\cite{queueingtheory} has derived the calculation formula of loss rate, that is $\frac{\rho^m - \rho^{m+1}}{1-\rho^{m+1}}$, where $\rho$ is the idle time of the queue. Therefore, we have:
$r^S_i = [(\rho_i)^{\theta S_i} - (\theta \rho_i)^{S_i+1}] / [ 1-(\rho_i)^{\theta S_i+1}]$, $r^L_i = [(\rho_i)^{L_i - \theta S_i} - (\rho_i)^{L_i - \theta S_i+1}] / [ 1-(\rho_i)^{L_i - \theta S_i+1}]$, then We can express the loss function in terms of known parameters and thresholds ($S_i, L_i$), after solve this optimization problem, we can get the optimal thresholds.

\end{document}